\documentclass[useAMS,usenatbib]{mn2e}
\usepackage{mathptmx}
\usepackage{graphicx}
\newcommand{\kms}{{\rm km\ s^{-1}}}
\def\HI{{\sc H\,i}}
\def\HII{{\sc H\,ii}}
\def\CO#1#2{$^{12}$CO ($J=#1{-}#2$)}
\title[Molecular gas and star formation in Arp~140]{The unusual distribution
  of molecular gas and star formation in Arp~140}

\author[H.~Cullen et al.]{H.~Cullen,$^1$ P.~Alexander,$^1$ D.~A.~Green,$^1$
  M.~Clemens$^2$ and K.~Sheth$^3$\\
$^1$Astrophysics Group, Cavendish Laboratory, 19 J.~J.~Thomson Ave.,
    Cambridge CB3 0HE\\
$^2$Osservatorio Astronomico, Vicolo dell'Osservatorio, 5, 35122 Padova, Italy\\
$^3$Spitzer Science Center, California Institute of Technology, Pasadena,
    CA 91125, USA}

\date{2006 September 6th}

\pagerange{\pageref{firstpage}--\pageref{lastpage}}
\pubyear{2006}
\begin{document}

\label{firstpage}

\maketitle

\begin{abstract}
We investigate the atomic and molecular interstellar medium and star formation
of NGC~275, the late-type spiral galaxy in Arp 140, which is interacting with
NGC~274, an early-type system. The atomic gas ({\HI}) observations reveal a
tidal tail from NGC~275 which extends many optical radii beyond the interacting
pair. The {\HI} morphology implies a prograde encounter between the galaxy pair
approximately $\sim 1.5 \times 10^{8}$ years ago. The H$\alpha$ emission from
NGC~275 indicates clumpy irregular star-formation, clumpiness which is mirrored
by the underlying mass distribution as traced by the $K_{\rm s}$-band emission.
The molecular gas distribution is striking in its anti-correlation with the
{\HII} regions. Despite the evolved nature of NGC~275's interaction and its
barred potential, neither the molecular gas nor the star formation are
centrally concentrated. We suggest that this structure results from stochastic
star formation leading to preferential consumption of the gas in certain
regions of the galaxy. In contrast to the often assumed picture of interacting
galaxies, NGC~275, which appears to be close to merger, does not display
enhanced or centrally concentrated star formation.
If the eventual merger is to lead to a significant burst of star
formation it must be preceded by a significant conversion of atomic to
molecular gas as at the current rate of star formation all the molecular gas
will be exhausted by the time the merger is complete.
\end{abstract}


\section{Introduction}

Despite the numerous studies associating galaxy interactions with enhanced star
formation \citep[e.g.,][]{1987AJ.....93.1011K, 2003MNRAS.346.1189L,
2004MNRAS.355..874N}, much progress is still to be made in the understanding of
the physics of this process. Whilst it is clear that interactions can lead to
enhanced star formation it is equally clear that they do not do so with
ubiquity \citep*{2003A&A...405...31B}.

In a study of interacting systems \cite{1987AJ.....93.1011K} find that star
formation in the nuclear region is more sensitive than the disk to the effects
of interactions. This observation was first reported by
\cite{1980A&A....89L...1H, 1981A&A....96..111H} in studies of radio emission
from interacting systems and was subsequently corroborated by the H$\alpha$
study of \cite{1986AJ.....91..255B}. More recently, \cite{2003A&A...405...31B}
observed a moderate increase in the star-formation rate (SFR) in the central
regions of interacting galaxies but found the global star formation properties
did not differ significantly from normal isolated systems.

Evidence that galaxy interactions have a more marked effect on the nuclear
rather than disk properties of galaxies is supported by $N$-body simulations
which suggest that interactions can drive a large fraction of gas into the
central regions of the galaxies \citep{1996ApJ...464..641M}. Nonetheless, the
distribution of star formation triggered in interacting galaxies is not clear
cut. Whilst many observations of interacting systems, including those of the
most extreme star formation found in Ultra Luminous Infrared Galaxies (ULIRGs),
indicate centrally concentrated star formation, observations over a range of
wavelengths reveal a number of interacting galaxies with more extended star
formation. The Antennae galaxies (NGC~4038/39) are an obvious example: despite
abundant dense gas in the two galactic centres, the most intense star formation
is found in the overlap region between the two galaxies
\citep{2004ApJS..154..193W}. Similarly, many interacting systems exhibit bright
spiral arm regions; a notable and well studied example is M51. Star formation
is also found in the very extended regions of interacting galaxies, for example
in tidal dwarfs \citep[e.g.,][]{2000AJ....120.1238D}, shells and large {\HII}
complexes in the outer regions as in M101 or NGC~628
\citep{2000AJ....120.1306L}.

Questions about the nature of star formation in interacting galaxies are linked
to those surrounding fuelling and the distribution of molecular gas; if star
formation is to occur there must exist an adequate and ongoing supply of gas.
Interactions can disrupt the axi-symmetry of the galactic potential leading to
gas inflow toward the central region. A number of studies have observed
enhanced and centrally condensed molecular gas emission in interacting systems,
although there is a bias towards more evolved interactions showing significant
perturbation, systems that are likely to be merger candidates
\citep{1988ApJ...334..613S, 1999IAUS..186..217Y, 2004A&A...422..941C}.

One important mode by which interactions can disrupt the galactic potential is
by the triggering of bar formation \citep*[e.g.,][]{1987MNRAS.228..635N,
1990ApJ...364..415E}. Simulations indicate bars can lead to gas flow towards
the central regions of galaxies \citep[e.g.,][]{1988A&A...203..259N,
1991A&A...243..118S, 1991ApJ...370L..65B}, results that are borne out by the
observational studies which reveal enhanced molecular gas emission in the
centres of barred systems \citep{1999ApJ...525..691S, 2005ApJ...632..217S}.
\cite{2005ApJ...632..217S}, who studied a sample of 50 nearby spirals --
including both late type barred and unbarred spirals, and CO-bright and
CO-faint galaxies -- found the effect of the bar equally pronounced in late and
early Hubble type spirals. They also note that in the case of some early-type
barred spirals there is little or no gas within the bar, which is consistent
with higher accretion rates provided these galaxies have undergone a starburst.
Understanding of the process of star formation in merge in local systems, which
can be studied in detail, is necessary for studies of star formation at higher
redshifts, where mergers are more frequent.

Here we undertake a detailed examination of the atomic and molecular gas in,
and star forming properties of NGC~275, the late-type, weakly barred spiral
(classified as SB(rs)cd pec) galaxy that is one component of the interacting,
spiral-lenticular pair Arp~140. The other component of Arp 140 is the
early-type system NGC~274 (classified as SAB(r)0$-$pec). This is one
of a small sample of nine spiral--elliptical interacting pairs, which have been
studied by \cite{2003Ap&SS.284..503C, 2006MNRAS.366...49C} and
\cite{2005dmu..conf..353C}. In undertaking these studies we hope to shed light
on the possible effects of interaction on late-type spiral systems and the
processes which may have shaped and will continue to shape NGC~275's evolution.
The recessional velocity of both NGC~275 and NGC~274 is 1750 $\kms$
\citep{1975ApJ...198..527S, 1991trcb.book.....D}, assuming an Hubble constant
of 70 km~s$^{-1}$ Mpc$^{-1}$, this corresponds to a distance of 25~Mpc, which
is used is this paper. The projected major axis diameter ($D_{25}$) for both
NGC~274 and NGC~275 is 10.9~kpc (1.5 arcmin) and the angular separation of the
pair is $44''$, giving a small projected linear separation of 5.3~kpc.

\section{Observations and data reduction}

\subsection{Optical observations}

Broadband optical $V$-band and narrowband H$\alpha$ images of Arp 140 were
obtained at the 2.3-m telescope at Siding Spring Observatory on the 2002
October 6th. The observations were taken with a $1024 \times 1024$ CCD with a
field of view of $6\farcm62$, and plate scale of 24.82 arcsec mm$^{-1}$. The
redshifted H$\alpha$ line of Arp 140 appears at 660.1 nm and was isolated using
a 5.0 nm filter centred on 658.5 nm. Six 5-minute exposures of Arp 140 were
observed at $10''$ offsets in the H$\alpha$ filter; short exposures were taken
to avoid tracking problems. A 60~s exposure was taken in $V$-band for continuum
subtraction. Exposures were made under photometric conditions and the spectral
standard star G158$-$100 was observed with both the narrowband and broadband
filters to allow calibration of the images. The images were bias and flat-field
corrected.  The individual narrowband images were then aligned, co-added and
the red continuum subtracted to produce the final image of 1800~s total
integration time and $\sim 1$ arcsec spatial resolution with a pixel scale of
$0\farcs599$ per pixel.

\subsection{21-cm observations}

\begin{table*}
\caption{Details of the VLA {\HI} observations.}\label{HI_obs}
\begin{tabular}{@{}cccc}\hline
Observing Array             & B-Array & C-Array & D-Array \\ \hline
Observing Dates             & 2003 Dec 10 and 19  & 2001 Jul 14      & 1987 May 21      \\
Pointing Centre: RA (J2000) &  00 51 04.20        &  00 51 03.00     &  00 51 01.95     \\
Dec (J2000)                 & $-$07 03 60.0       & $-$07 03 30.0    & $-$07 03 23.2    \\
Channels                    & 127                 &  63              & 31               \\
Velocity Width ($\kms$)     & 5.2                 & 20.7             & 42.6             \\
Time On-source (minutes)    & 447                 & 113              & 58               \\
Flux Calibrator             & 3C48 (15.95~Jy)     & 3C48 (15.95~Jy)  & 3C48 (15.95\~Jy) \\
Phase Calibrator            & 0059$+$001          & 0054$-$035       & 0056$-$001       \\ \hline
Angular Resolution          & $7\farcs24 \times 5\farcs08$  & $23\farcs9 \times 20\farcs7$   & $113\farcs5 \times 49\farcs9$  \\
of Channel Maps             &                               &                                &                                \\
Channels Used for           & $2{-}5$ and $104{-}114$       & $2{-}16$ and $51{-}61$         & $1{-}9$ and $24{-}30$          \\
Continuum Subtraction       &                               &                                &                                \\ \hline
\end{tabular}
\end{table*}

VLA {\HI} observations were in both B-, C- and D-array data were retrieved from
the National Radio Astronomy Observatory (NRAO) archive. Details of all of the
{\HI} observations are presented in Table~\ref{HI_obs}. The data reduction
followed standard procedures, using NRAOs Astronomical Image Processing System
({\sc AIPS}) package. Channel maps were made, details of which are given in
Table~\ref{HI_obs}. For the C- and D-array data combined a $uv$-taper of
9~k$\lambda$ was applied (in $u$) to help correct for the distorted beam, a
result of the low declination of the source. Continuum emission was removed by
subtraction of the average of the line-free channels. Channel maps were summed
in MOMNT using a combination of `cut-off' and `window' methods
\citep*{1999sira.conf.....T}, to optimise detection of the {\HI} emission.

Line-free channels from the VLA B-array data were used to produce an L-band
continuum map of NGC~275. A conservative choice of channels $104{-}114$
($1484{-}1536$ $\kms$) in the 128 channel-wide bandwidth was selected to sample
the continuum emission. This limited channel range excluded all potential line
emission and the edge-of-bandwidth channels displaying interference. The data
were self-calibrated and imaged using natural weighting to increase the
sensitivity to large-scale structure. The flux density obtained ($72 \pm
5$~mJy) is in good agreement with the NVSS flux density (60~mJy) for NGC~275
\citep{1998AJ....115.1693C}.

\subsection{\CO{1}{0} observations}

Observations of NGC~275 were made using both the six-element Owens Valley Radio
Observatory (OVRO) millimeter interferometer and the 10-element
Berkeley-Illinois-Maryland Association (BIMA) millimeter interferometer. In
both cases the \CO{1}{0} line was observed at 115.2712 GHz. In total, three
tracks were observed using the BIMA interferometer in C-configuration on 2003
March 23, 26 and May 18, but the data taken on March 26 were unusable because
of high levels of interference. The BIMA correlator was configured to have a
resolution of 1.56~MHz ($4~\kms$) over a total bandwidth of 367~MHz
($955~\kms$). The primary beam of BIMA at this frequency is $100''$ FWHM. The
sources 3C454.3 and 0108$+$015 were observed for passband and phase calibration
respectively. System temperatures averaged $\sim550$~K for the March
observations and $\sim400$~K for the May observations. In addition, NGC~275 was
observed using the OVRO interferometer in both C- and L-configurations on 2003
October 3 and 31. The primary beam at OVRO at 115 GHz is $66''$. 3C84 and
3C454.3 were used as passband calibrators and J0116$-$116 was observed every 20
minutes for phase calibration. Typical single-sideband system temperatures
ranged between 450 and 700~K in C-configuration and between 300 and 450~K in L.
The OVRO correlator was configured with 120 independent channels of 2 MHz
($5.2~\kms$) giving a total bandwidth of 240~MHz ($625~\kms$).

Data calibration was carried out in {\sc MMA} \citep{1993PASP..105.1482S} and
{\sc Miriad} \citep{1995ASPC...77..433S} for the OVRO and BIMA datasets
respectively. Calibrated datasets were then read into {\sc AIPS}, further
flagging applied and the calibrated $uv$ datasets combined. Data was binned to
a spectral resolution of $4~\kms$. A $uv$-taper of 20~k$\lambda$ was applied in
an attempt to reduce the highly distorted beam resulting from the low
declination of the source. The resulting resolution of the channel maps was
$10\farcs30 \times 7\farcs63$.

\subsection{\CO{2}{1} observations}

Observations of the \CO{2}{1} line at 230.54 GHz towards NGC~275 were made with
the James Clerk Maxwell Telescope (JCMT) in `service' mode between 2002 August
19 and September 20. Observations were made on an $8 \times 8$ grid with $10''$
spacing in both RA and Dec, centred on RA $00^{\rm h}$ $51^{\rm m}$ $04\fs29$,
Dec $-07^\circ$ $03'$ $56\farcs6$ (J2000). The pointing separation of $10''$
produced a fully sampled map (the JCMT beam is $20\farcs8$ at 230.54 GHz). The
observing strategy adopted sampled the whole grid at 1 minute per point. This
was repeated 14 times to ensure comparable noise levels across the map.  Sky
subtraction was achieved by beam-switching to a point $120''$ in RA from the
reference position. Scans at each position were averaged. Each spectrum was
corrected for a residual frequency-dependent baseline offset by removing a
linear baseline fitted to end channels. The spectra were then binned to a
velocity resolution of $20.4~\kms$.

Standard calibration observations were made at the same time as the
observations of NGC~275. Data calibration followed the standard JCMT procedure,
adopting the beam efficiency and forward scattering efficiency on the JCMT
website yielding an $\eta_{\rm fss}$ of 0.77. To convert between main beam
temperature and flux for NGC~275, it was assumed that the \CO{2}{1} emission
could be modelled as a point source. Whilst this is not strictly the case, it
allows us to place a lower limit on the \CO{2}{1} flux for this system. The
conversion between flux and main beam temperature assuming a point source is
given by $S = 18.4T_{\rm mb}$, where we take the FWHM of the JCMT beam at
230~GHz to be $20\farcs8$.

\section{Results}

\subsection{Neutral atomic hydrogen in Arp 140}

\begin{figure*}
\centerline{\includegraphics[clip=,angle=90,width=15cm]{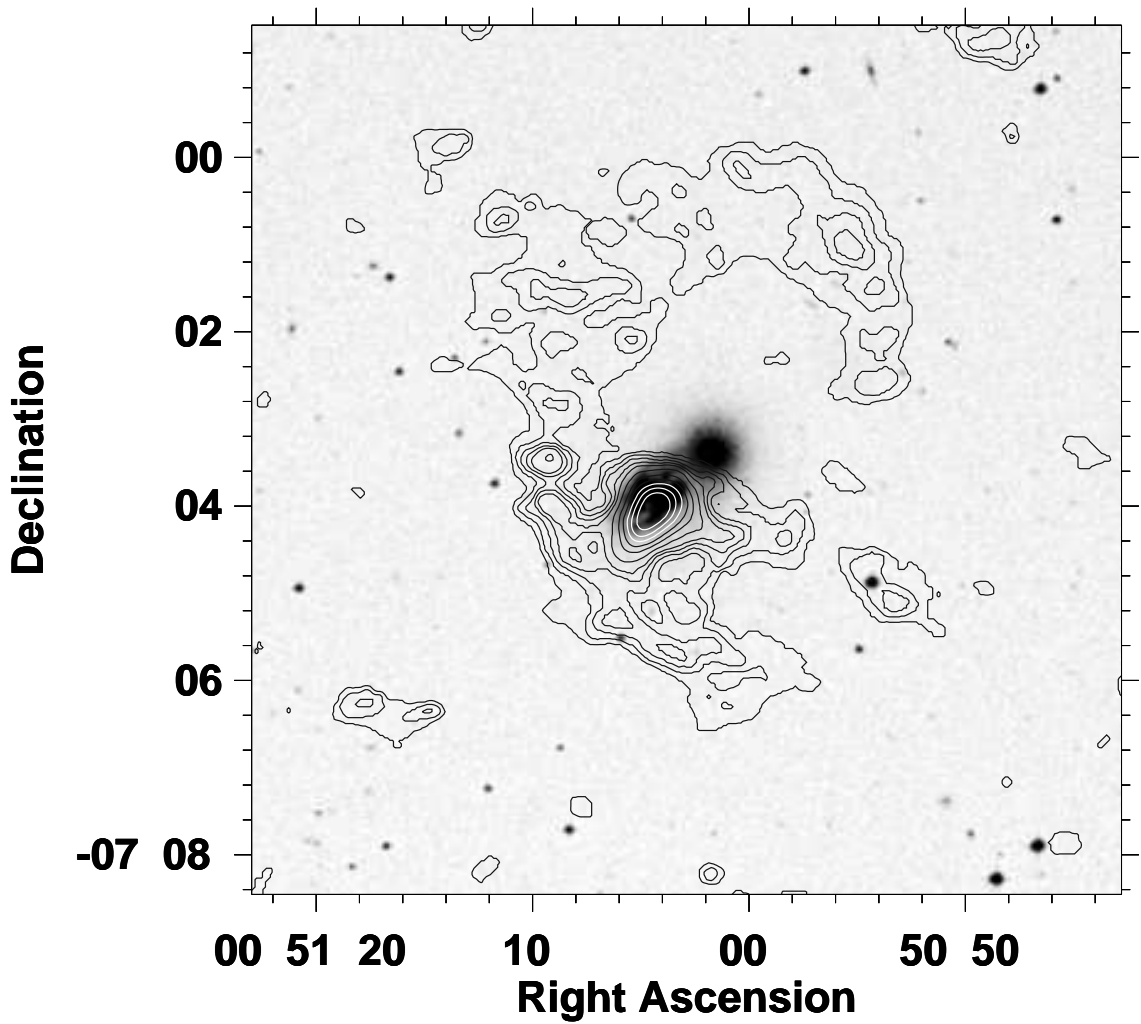}}
\caption{Integrated {\HI} emission from Arp 140, from VLA C- and D-array
observations ($\theta_{\rm FWHM} = 23\farcs88 \times 20\farcs70$), overlayed on
a digitised sky survey $R$-band image. Contour levels are (0.03, 0.09, 0.15,
0.21, 0.27, 0.4, $0.6\dots$ 1.6) Jy~beam$^{-1}$ km~s$^{-1}$.}
\label{a140_mom0c.fig}
\end{figure*}

\begin{figure*}
\centerline{\includegraphics[clip=,angle=90,width=15cm]{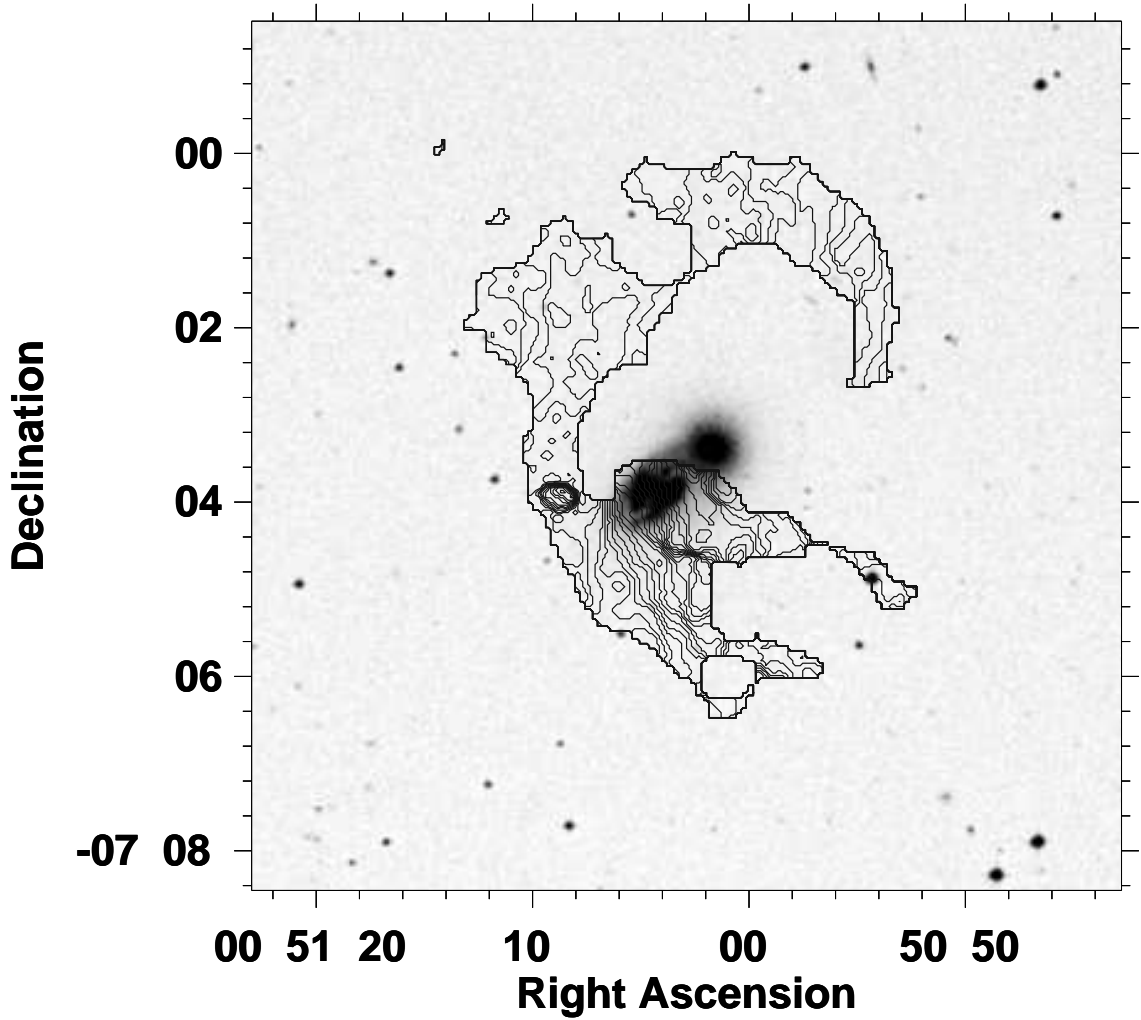}}
\caption{Intensity weighted mean velocity field of Arp 140 using the VLA
C- and D-array data. Contour levels are (1580, 1590,
1600$\ldots$ 1940) $\kms$}\label{a140_hivel.fig}
\end{figure*}

\begin{figure*}
\centerline{\includegraphics[clip=,angle=90,width=12cm]{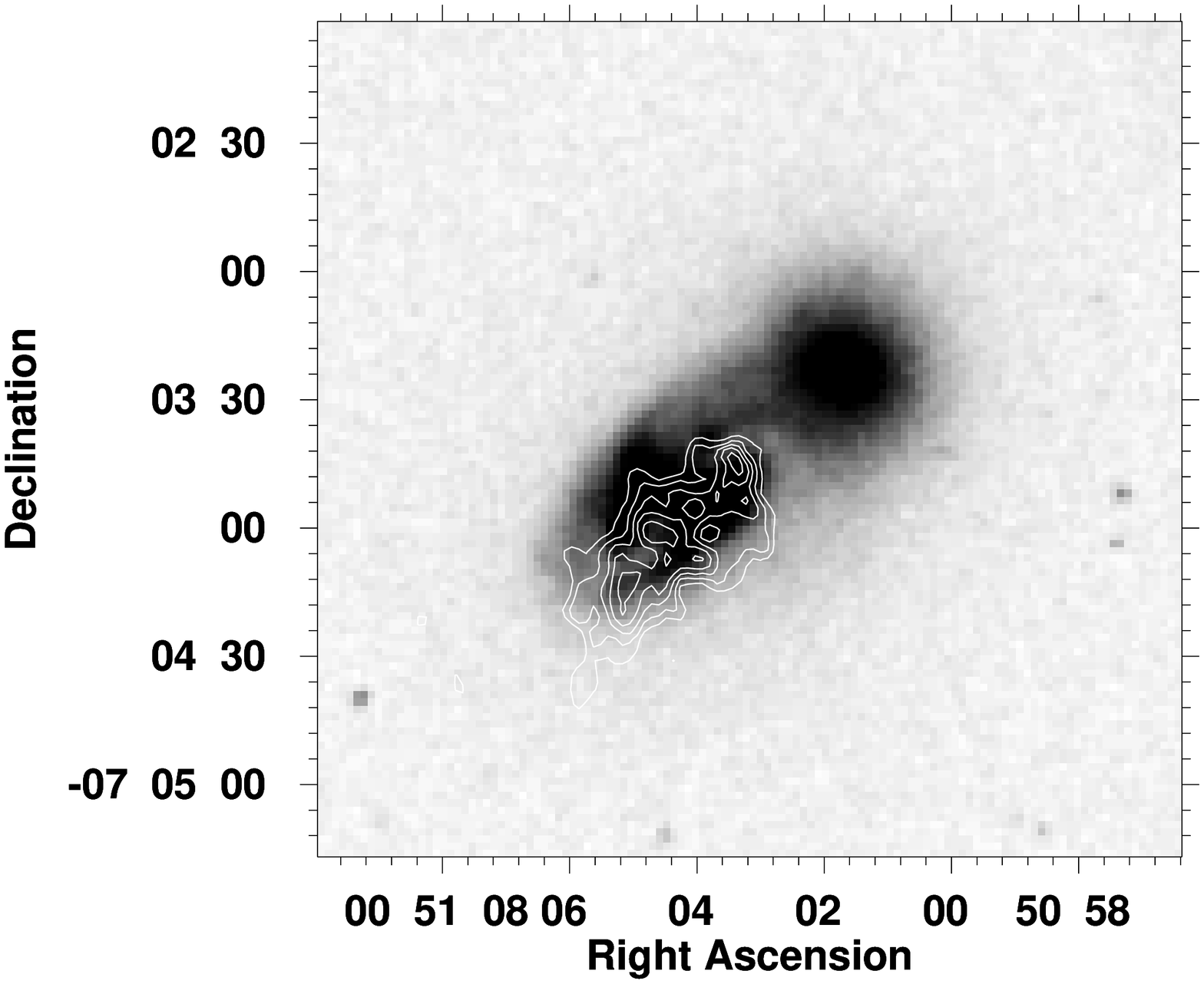}}
\medskip
\centerline{\includegraphics[clip=,angle=90,width=12cm]{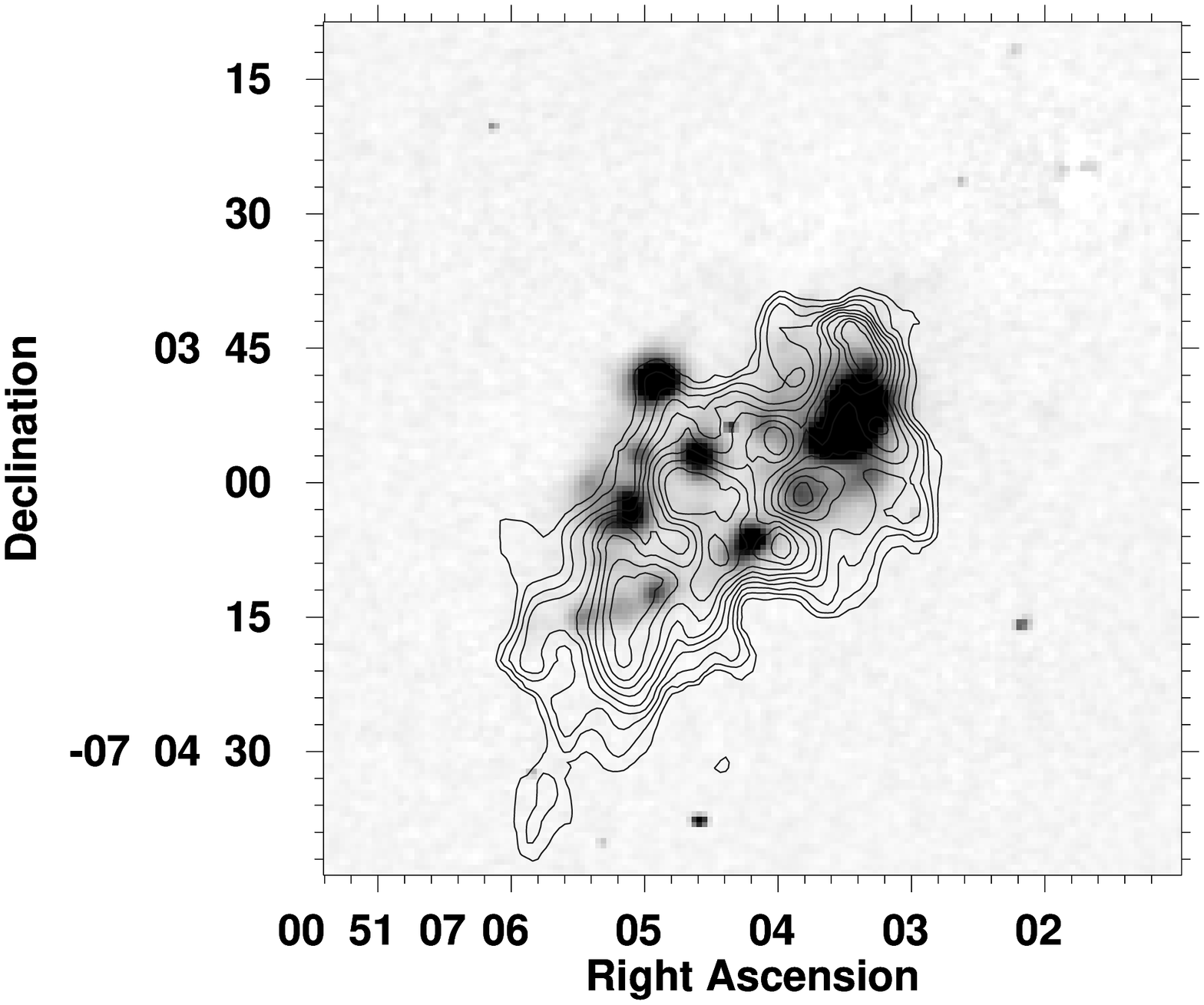}}
\caption{Integrated {\HI} emission from NGC~275, from  VLA B-array observations
($\theta_{\rm FWHM} = 7\farcs24\times 5\farcs08$), overlayed a digitised sky
survey $R$-band image (top) and a continuum subtracted H$\alpha$ image
(bottom). Contour levels are (0.06, 0.1, 0.14, 0.18, 0.22, 0.26) Jy~beam$^{-1}$
km~s$^{-1}$.}\label{a140_mom0b.fig}
\end{figure*}

\begin{figure*}
\centerline{\includegraphics[clip=,angle=90,width=8cm]{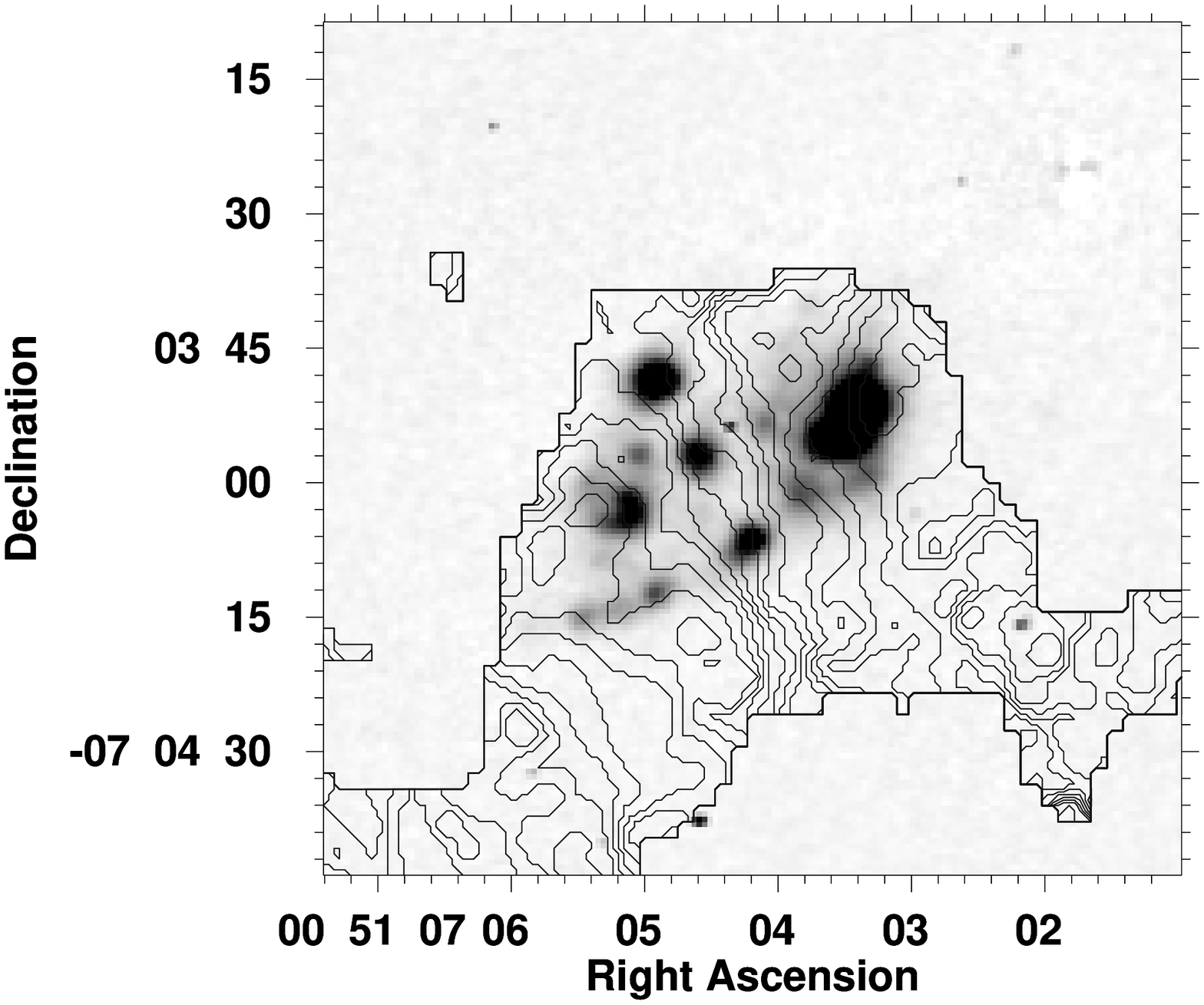}
            \qquad
            \includegraphics[clip=,angle=90,width=8cm]{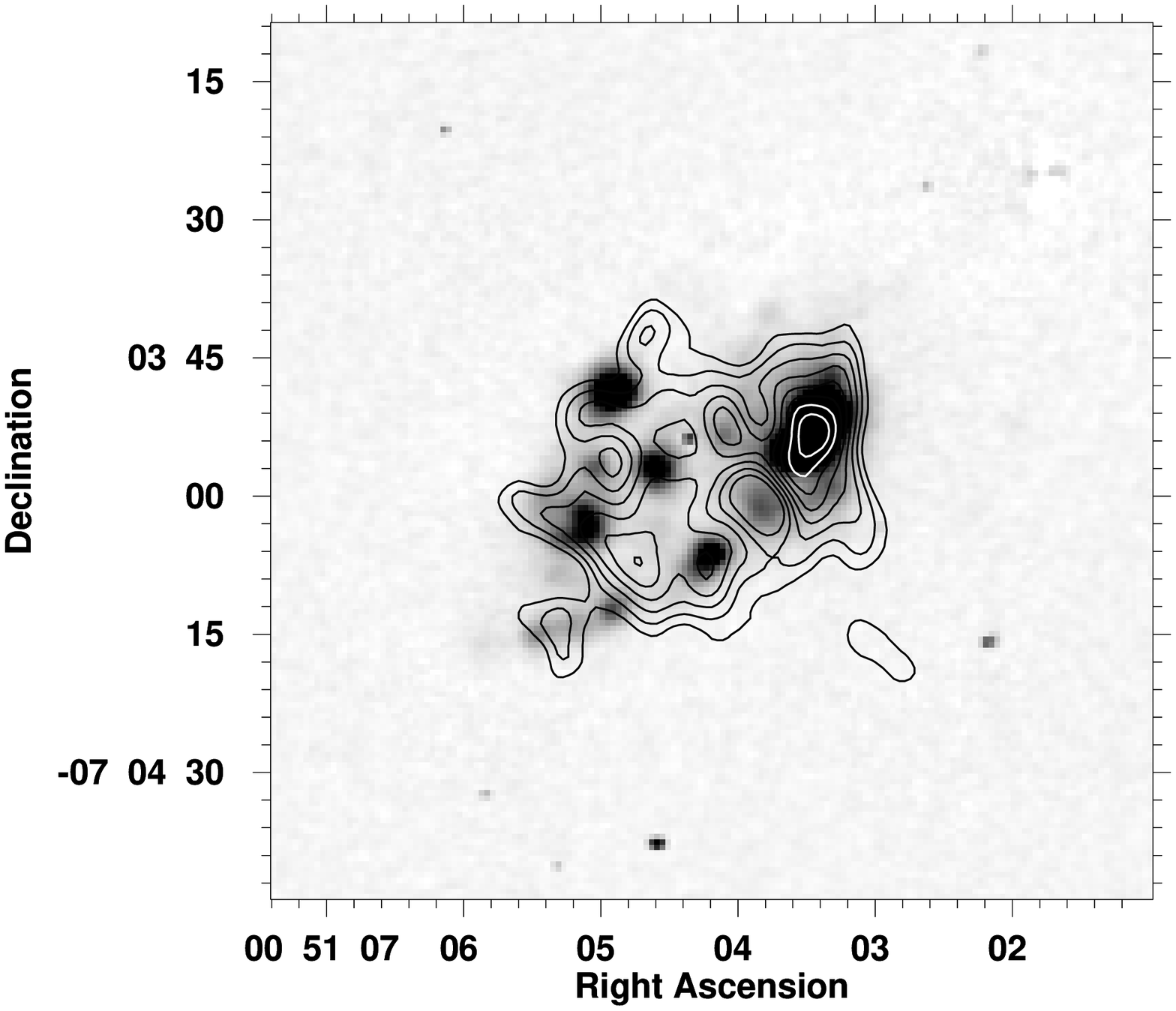}}
\caption{(left) Intensity weighted mean velocity field of NGC~275 naturally
weighted combined VLA B and C array data ($\theta_{\rm FWHM} = 8\farcs64\times
6\farcs17$) for NGC~275 overlayed on a continuum subtracted H$\alpha$ image.
Contour levels correspond to (1650, 1660, 1670$\ldots$1940)$\kms$. (right) VLA
B Array ($\theta_{\rm FWHM} = 7\farcs33 \times 5\farcs07$) radio continuum
emission of NGC~275 overlayed on continuum subtracted H$\alpha$ emission.
Contours at: 1.2, 1.5, 1.8, 2.1, 2.4, 2.7, 3.0, 3.3, 3.6 mJy
beam$^{-1}$.}\label{275_hivel_ha_op.fig}
\end{figure*}

The VLA C- and D-array {\HI} map is shown in Fig.~\ref{a140_mom0c.fig},
together with the velocity field shown in Fig.~\ref{a140_hivel.fig}. There is
an extended tidal tail associated with the Arp 140 system.
Given the galaxies were selected as a late-type interacting
with a gas-poor system, ram pressure from the cool ISM of the later is
not expected to be important.
The distribution of gas is consistent with the {\HI} originating in NGC~275.
The peak flux is 1.7 Jy beam$^{-1}$ km s$^{-1}$ corresponding to a column
density of $3.8 \times 10^{21}$ cm$^{-2}$ ($\sim 31$ M$_{\odot}$ parsec$^{-2}$)
falling to around 0.1 Jy beam$^{-1}$ km s$^{-1}$, corresponding to a column
density of $2.3 \times 10^{20}$ cm$^{-2}$ ($\sim 2$ M$_{\odot}$ parsec$^{-2}$)
in the extended regions of the tidal tail. Two peaks are observed to the east
of the optical system near $00^{\rm h}$ $51^{\rm m}$ $09^{\rm s}$, at
declinations $-07^\circ$ $03'$ $54''$ and $-07^\circ$ $03'$ $22''$; both have
velocities considerably larger than those of the surrounding {\HI}. Both
regions also exhibit large velocity dispersion, of order $100~\kms$, in
contrast to the average dispersion of about $15~\kms$ observed in much of the
extended southern tidal tail. Excluding the two peaks described above -- as
they may not be related to the tidal tail -- the peak flux observed in the
tidal tail is 0.25 Jy beam$^{-1}$ km s$^{-1}$, a column density of $5.6 \times
10^{20}$ cm$^{-2}$ ($\sim 4.5$ M$_{\odot}$ parsec$^{-2}$).

The total {\HI} flux for Arp 140, obtained by adding the flux in the
low-resolution channel maps, is 25.9~Jy km~s$^{-1}$, which is in agreement with
the single dish flux of $28.5 \pm 4.3$ Jy km~s$^{-1}$
\citep{1975ApJ...198..527S}. This corresponds to a neutral hydrogen mass of
$(3.8 \pm 0.4) \times 10^{9}$ M$_{\odot}$. The integrated {\HI} map in
Fig.~\ref{a140_mom0c.fig} reveals no emission from the early-type system,
NGC~274.  Based on an rms sensitivity of $5 \times 10^{-4}$ Jy beam$^{-1}$, the
upper limit on the {\HI} content of NGC~274 of $4.4\times 10^{6}$ M$_{\odot}$
(0.03 Jy km~s$^{-1}$).

The southern region of the tidal arm has a velocity dispersion of $\sim
200~\kms$ with velocities ranging from $1583~\kms$ on the eastern side of the
tail to $1779~\kms$ on the western side. The largest blue shifted point in the
{\HI} emission ($1583~\kms$) is located at the far eastern edge of the
northerly extension of the tidal tail, whereas the largest redshifted point
($1917~\kms$) is found on the most western point of the southerly extension of
the tail.  The more northern of the two {\HI} peaks observed at the base of the
northern tidal tail in Fig.~\ref{a140_mom0c.fig} is associated with an abrupt
change in the velocity profile of the {\HI} emission.  Examining the spectra
over this region reveals a $20'' \times 24''$ ($\sim 1$ beam) area displaying
an {\HI} profile with two peaks.  This is consistent with emission from
stripped gas associated with Arp 140 at $1646 \pm 20$ km~s$^{-1}$ and a small
{\HI} cloud (at $1854 \pm 20$ km~s$^{-1}$), with a mass in excess of $9 \times
10^{6}$ M$_{\odot}$, which may be tidal debris.

We now consider the likely origin of the extended {\HI} structure we observe.
This stucture is consistent with tidal stripping, and we feel that it is
unlikely to have been significantly influenced by ram-pressure effects.
NGC 274 and 275 are members of a poor group which also includes NGC 298 and 337
(see Hyperleda), not a rich cluster where a significant intra-cluster medium
may be present and ram-pressure effects could be important.
NGC 274 is a gas-poor early-type galaxy with little evidence of any {\HI}
being associated with it: the northern tidal tail implies that the pair rotate
in an anti-clockwise sense as projected in Fig.~1. This further suggests that
the southern displacement of atomic gas relative to the centre of NGC 275 is
not caused by ram-pressure from an extended medium. In this case the
displacement should be in the opposite sense.

\subsection{High resolution {\HI} data}

A high-resolution ($\theta_{\rm FWHM} = 7\farcs24 \times 5\farcs08$) map made
using B-array data is shown in Fig.~\ref{a140_mom0b.fig}; the angular
resolution of these data corresponds to a physical scale of $\sim 0.7$ kpc.
Whilst the major axis of the {\HI} emission in NGC~275 is parallel to that of
the $R$-band optical data, it lies approximately $\sim 7''$ ($\sim 0.8$~kpc) to
the south west of the optical major axis. The atomic gas associated with the
galaxy has a clumpy distribution with three main peaks aligned approximately
along the major axis. The northern and southern peaks are offset from the
central peak by $23''$ (2.8~kpc) and $16''$ (2~kpc) respectively. The clear
offset of the atomic gas with respect to the stellar structure is striking.
Similar displacements of the {\HI} emission from the stellar disks were
observed by \cite{2005ApJS..158....1I} in their study of ten interacting
systems.

Figure~\ref{275_hivel_ha_op.fig} shows the intensity-weighted mean velocity
field of NGC~275 using the combined VLA B- and C-array data. The velocity
contours indicate increasingly blueshifted gas as we move along the major axis
of NGC~275 from south-west to north-east. This progression continues to the
base of the tidal arm extending northward from the galaxy, where a number of
loops in the contour lines are observed indicative of a disrupted velocity
field. The velocities observed in this region are $\sim 1700$ $\kms$, and we
identify this region as the transition between atomic gas still associated with
the galaxy and gas in the tidal tail. This yields a velocity range of gas
associated with NGC~275 of $1697{-}1929$ $\kms$.

The structure observed in the B-array map therefore predominantly traces atomic
gas still bound to the galaxy. The integrated {\HI} flux for this bound gas is
$8.9 \pm 0.9$ Jy $\kms$ (linewidth $313~\kms$), which corresponds to an {\HI}
mass of $(1.3\pm0.1) \times 10^{9}$ M$_{\odot}$, which is approximately one
third of the total {\HI} in the Arp~140 system. This is consistent with the
trend found by \cite{1996AJ....111..655H} who find for a sample of galaxies
which lie in the central region of the merger sequence that $\sim35\%$ of the
total atomic gas emission comes from the inner regions of the system, with this
percentage falling dramatically for more advanced mergers.

\begin{figure}
\centerline{\includegraphics[clip=,angle=270,width=8.5cm]{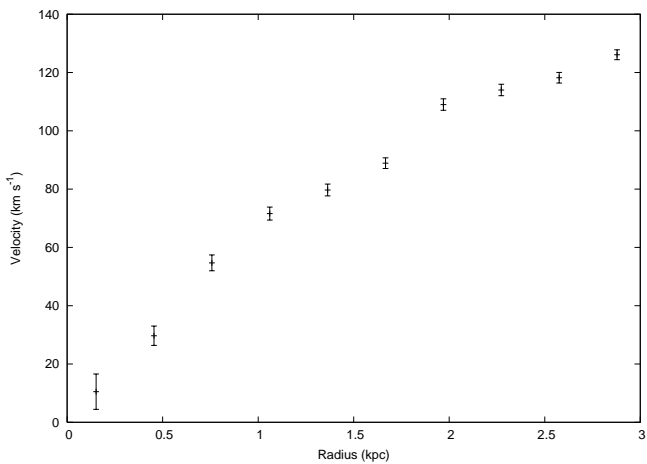}}
\caption{The model circular velocity de-projected on the galaxy plane as
obtained by fitting a tilted-ring model to the VLA B-Array
data.}\label{n275_rotcurv.fig}
\end{figure}

We have derived the rotation curve for NGC~275 (Fig.~\ref{n275_rotcurv.fig})
from the intensity-weighted velocity field using the tilted ring model
\citep*{1973MNRAS.163..163W} in the {\sc AIPS} task GAL. This method
approximates the galaxy as a set of concentric rings of gas at increasing
radii, characterised by a circular velocity, position angle and inclination
angle. The dynamical centre was fixed using the 2MASS $K_{\rm s}$-band data and
the inclination angle of 36.8 degrees was taken from the Hyperleda
database\footnote{See {\tt http://leda.univ-lyon1.fr/}}. The {\HI} rotation
curve was fitted by breaking the galaxy into annuli $2\farcs5$ in width (i.e.\
approximately a half-beam width) and fitting the velocity field in each
annulus, keeping the dynamical centre and inclination angle fixed.

\subsection{Star-formation in NGC~275}\label{275_star_form_rate.sec}

The calibrated H$\alpha$ data for the late-type system, NGC~275, yields a total
integrated flux of approximately $\sim 1.3 \times 10^{-15}$ W~m$^{-2}$, which
is equivalent to a luminosity of $9.7 \times 10^{33}$ W. Assuming no reddening
and adopting the calibration of \cite{1998ARA&A..36..189K} this corresponds to
a SFR of 0.77 M$_{\odot}$ yr$^{-1}$, which is in good agreement with the value
obtained by \cite{1989ApJ...345..176L} who quote a SFR for NGC~275 of 0.75
M$_{\odot}$ yr$^{-1}$. The far-infrared (FIR) emission gives an
obscuration-free estimate of the SFR: adopting the calibration of
\cite{2003ApJ...586..794B} we obtain a FIR SFR of 1.5 M$_{\odot}$ yr$^{-1}$ for
NGC~275, which we adopt for the remainder of this paper.

\begin{figure}
\centerline{\includegraphics[clip=,angle=90,width=8cm]{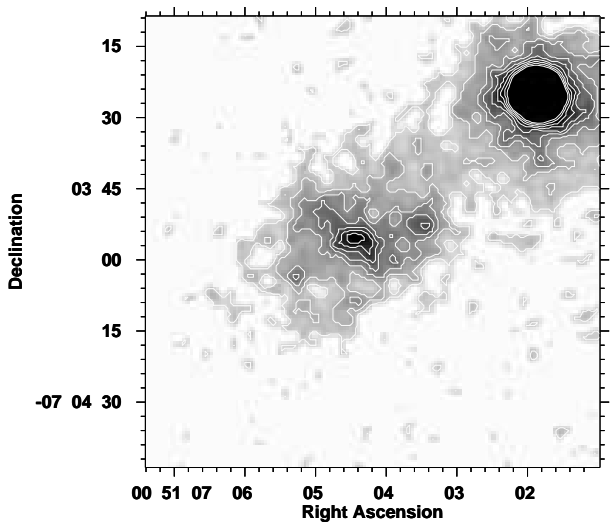}}
\caption{2MASS $K_{\rm s}$-band image of Arp 140.}\label{ngc275_ksml.fig}
\end{figure}

The star formation in NGC~275 as traced both by the H$\alpha$ and radio
continuum emission is irregular (Fig.~\ref{275_hivel_ha_op.fig}),
and is not centrally concentrated. Interestingly, the patchy star
formation traced by the H$\alpha$ emission is mirrored by the 2MASS $K_{\rm
s}$-band image which is shown in Fig.~\ref{ngc275_ksml.fig}. The near infrared
is usually considered a good tracer of the older stellar population and hence
of the mass distribution.  One caveat to this assumption is the possible
contribution to the near infrared from red supergiants, however this is
unlikely in NGC~275 given the observed star formation rate. The irregular
structure observed in the $K_{\rm s}$-band image is consistent with an
irregular underlying mass distribution. Fig.~\ref{ngc275_ksml.fig} shows a bar
at a position angle of about $45^\circ$, with two knots of H$\alpha$ at a
similar angle, but offset slightly to the leading edge. Such an offset has been
seen in previous studies \cite[see][for a study of six barred
spirals]{2002AJ....124.2581S}, where {\HII} regions were offset to the leading
edge of CO bars (see also \cite{2000ApJ...532..221S}). Also, enhanced star
formation was see at the ends of bars, as is indicated in the case of NGC 275
from H$\alpha$ knots nears the ends of the bar.

\subsection{Molecular gas}\label{ngc275_co.sec}

\subsubsection{Distribution of \CO{1}{0}}

\begin{figure*}
\centerline{\includegraphics[clip=,angle=90,width=8cm]{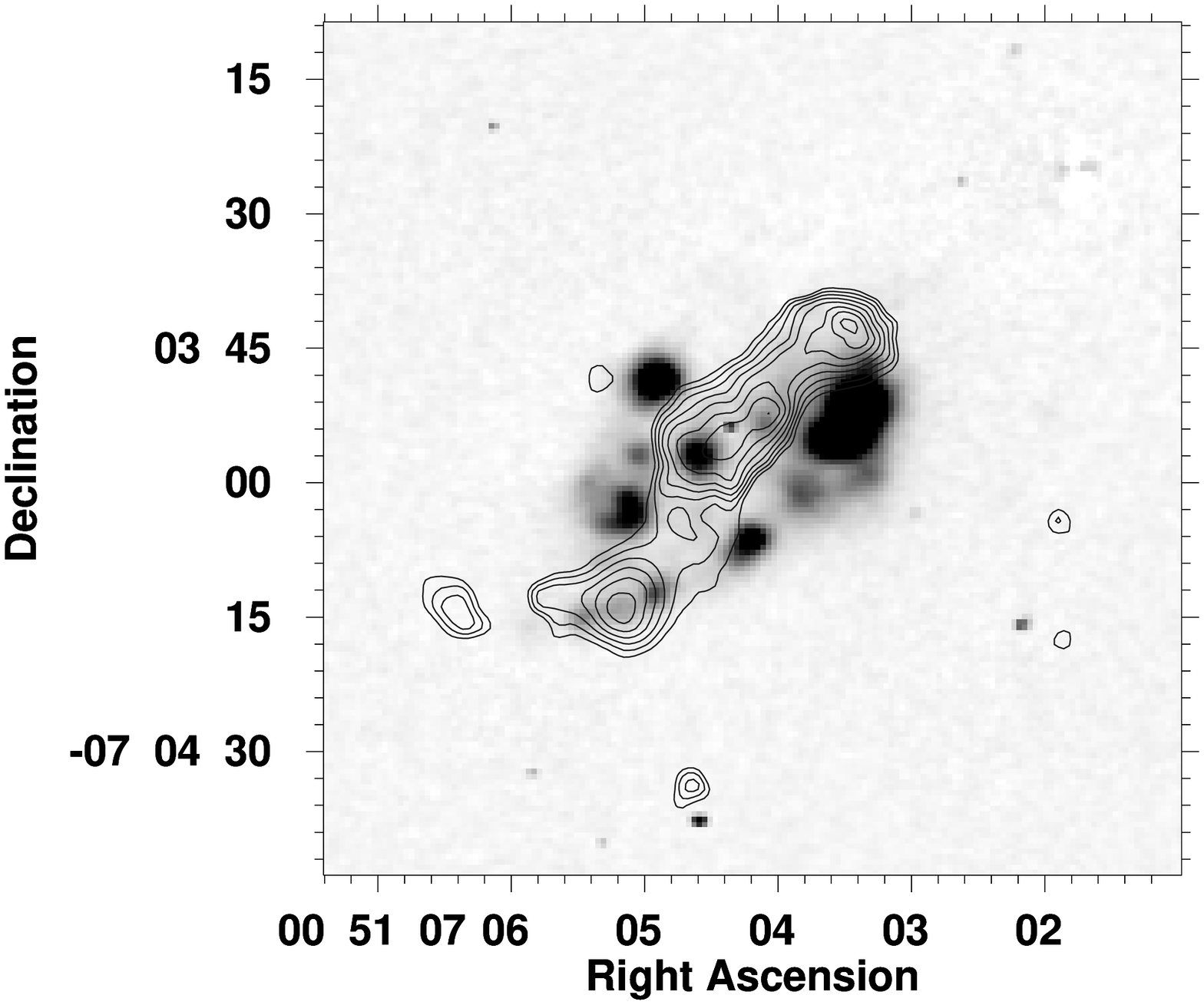}
            \qquad
            \includegraphics[clip=,angle=90,width=8cm]{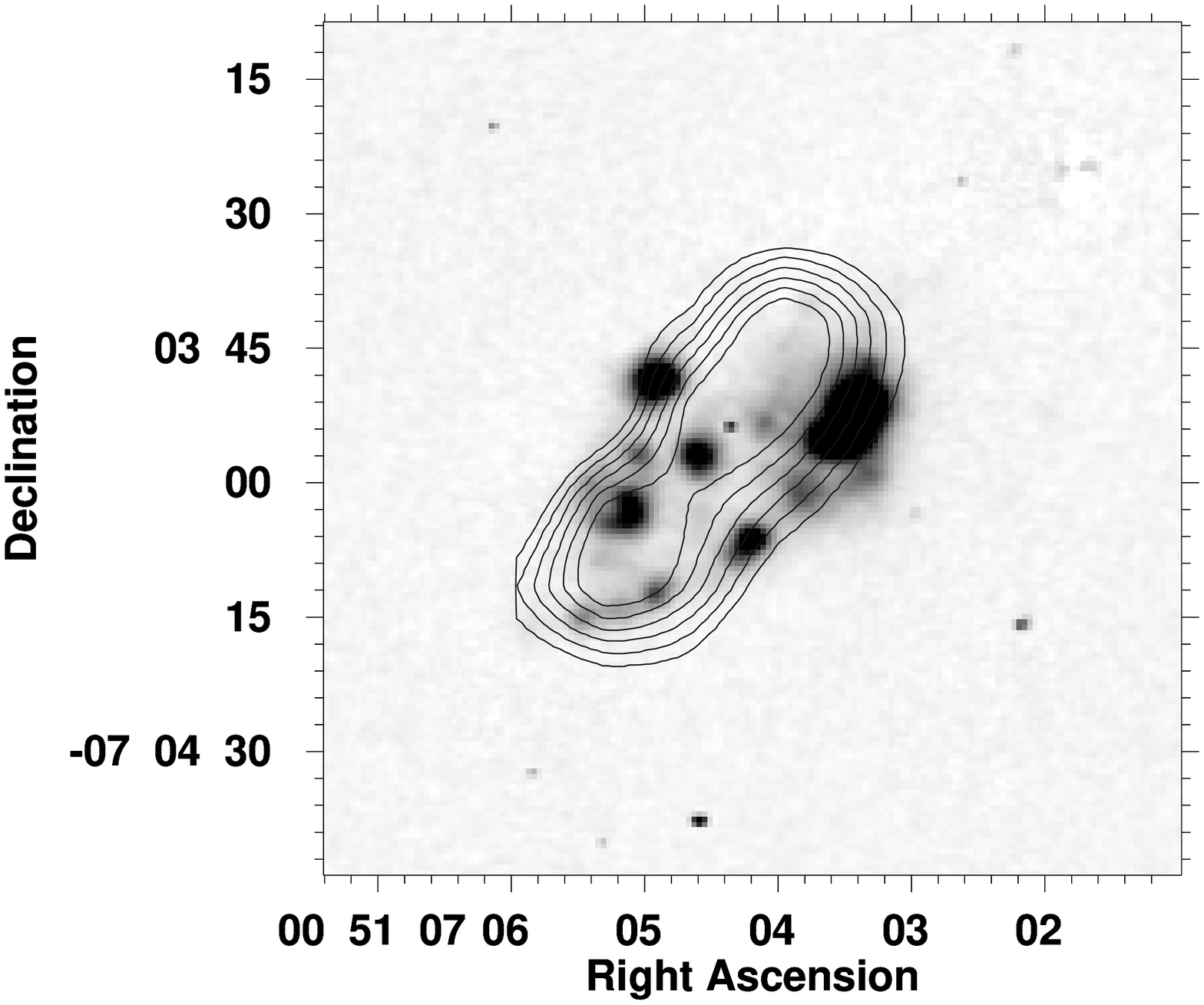}}
\caption{(left) \CO{1}{0} combined BIMA and OVRO data ($\theta_{\rm FWHM} =
10\farcs30 \times 7\farcs63$) for NGC~275. Contour levels correspond to (2.0,
2.5, 3.0, 3.5, 4.0, 4.5, 5.0, 5.5) Jy~beam$^{-1}$ km~s$^{-1}$. Data overlayed
on H$\alpha$ continuum subtracted image. {\bf Right:} JCMT \CO{2}{1} for
NGC~275 ($\theta_{\rm FWHM} = 22''$), contour levels are (1.8, 2.1, 2.4, 2.7,
3.0, 3.3, 3.6) K~km~s$^{-1}$.}\label{a140_comom0ha.fig}
\end{figure*}

A \CO{1}{0} map from the combined BIMA and OVRO data is shown in
Fig.~\ref{a140_comom0ha.fig}; the synthesised beam has  $\theta_{\rm FWHM} =
10\farcs30 \times 7\farcs63$, corresponding to a physical scale of $\sim
1$~kpc. The integrated emission is shown overlayed on a continuum subtracted
H$\alpha$ image (Fig.~\ref{a140_comom0ha.fig}). The CO data reveal three clear
peaks, of which the central peak is the most extended. Unlike the high
resolution {\HI} data the molecular gas appears well aligned with the major
axis of the optical $R$-band image. The integrated \CO{1}{0} flux is $110 \pm
20$ Jy $\kms$; adopting the CO-to-H$_{2}$ conversion factor $X \equiv
N(H_{2})/I_{CO} = 2.8 \times10^{20}$ cm$^{-2}$ K km s$^{-1}$
\citep{1986A&A...154...25B}, this yields a total mass of H$_{2}$ of $(7.5 \pm
1.5) \times 10^{8}$ M$_{\odot}$.

The velocity gradient of the molecular gas measured from the high resolution
\CO{1}{0} data yields a relatively constant velocity gradient (correspond to a
velocity change of $32.1 \pm 2.7~\kms$ over the central 3.6 kpc), in good
agreement with the value obtained from the {\HI} data ($33.2 \pm 0.7 \kms$).

\subsubsection{Distribution of \CO{2}{1}}

The \CO{2}{1} emission observed with the JCMT is shown in
Fig.~\ref{a140_comom0ha.fig}. CO emission is detected over a region
approximately $70'' \times 40''$, coincident with optical emission. The
observed \CO{2}{1} emission is consistent with the structure in the \CO{1}{0}
emission after its has been convolved to a similar resolution ($\theta_{\rm
FWHM}\sim 20''$). The total \CO{2}{1} flux for NGC~275 is 13.4 K$~\kms$, which
corresponds to a flux of $\sim 250$ Jy~km~s$^{-1}$ for the JCMT at 230~GHz.

\section{The interstellar medium in Arp 140}

The molecular gas mass of NGC~275 is $(7.5 \pm 1.5) \times 10^{8}$ M$_{\odot}$,
while the atomic gas mass  we estimate to be bound to NGC~275 is $(1.3 \pm 0.1)
\times 10^{9}$ M$_{\odot}$: the total atomic gas mass associated with the Arp
140 system is $(3.8 \pm 0.4) \times 10^{9}$ M$_{\odot}$. After allowing for
differences in the H$_{2}$ to CO conversion factor ($X \equiv N(H_{2})/I_{CO}$)
the molecular gas mass normalised by either the galaxy area or the $B$-band
luminosity is typical of other Scd galaxies observed by
\cite{1989ApJ...347L..55Y}, although it is towards the higher end of the range
obtained by \cite{1990A&A...233..357C} and \cite*{2003A&A...405....5B}.  The
atomic gas mass is also comparable to other Scd galaxies. Compared to other
interacting galaxies \citep*{2004A&A...422..941C} Arp 140 has a slight deficit
of molecular gas and a typical atomic gas mass.

\begin{figure}
\centerline{\includegraphics[clip=,angle=90,width=85mm]{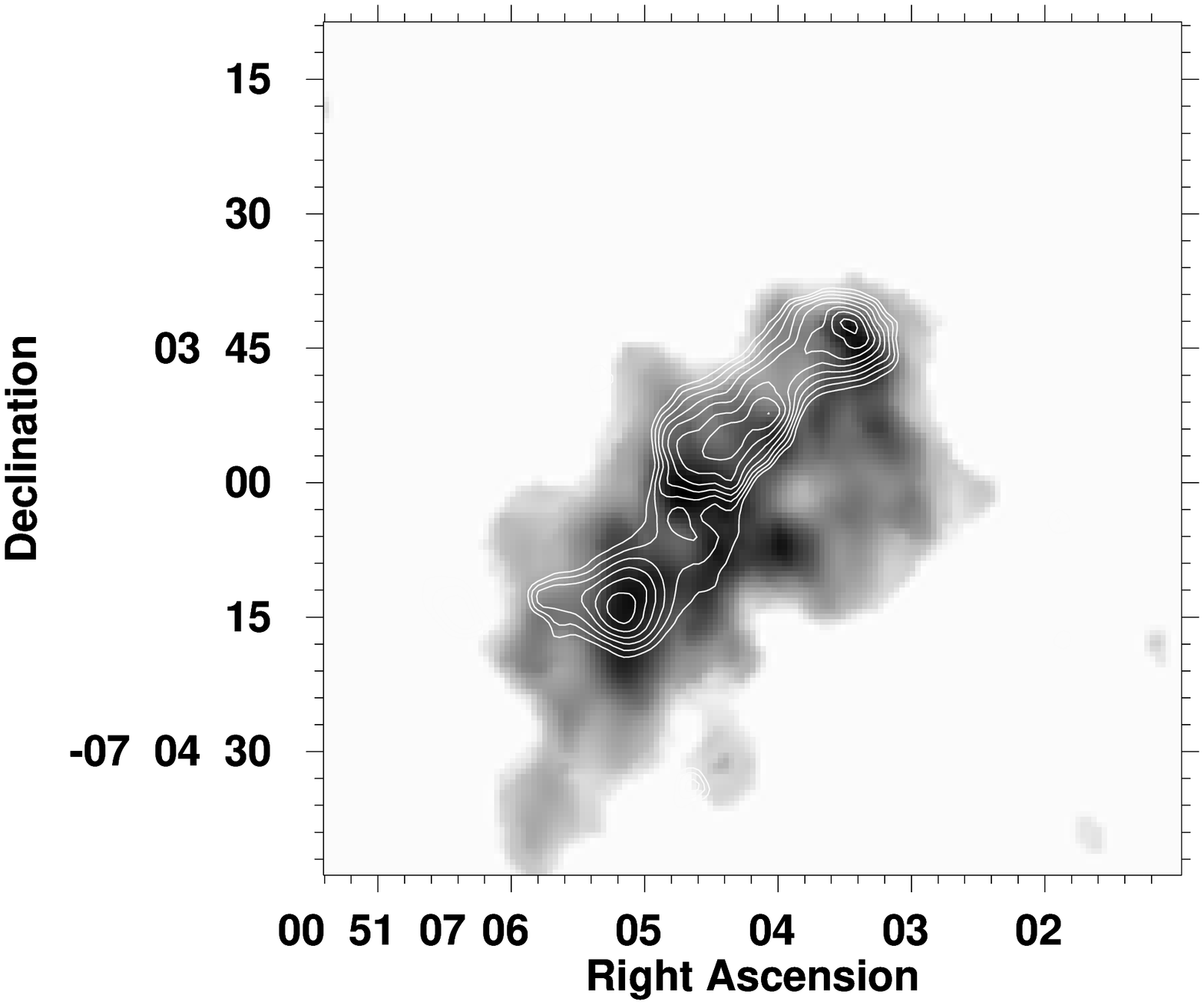}}
\caption{Contours of \CO{1}{0} emission from combined BIMA and OVRO data
($\theta_{\rm FWHM} = 10\farcs30 \times 7\farcs63$), overlayed on greyscale of
integrated {\HI} emission from VLA B-array observations ($\theta_{\rm FWHM} =
7\farcs24 \times 5\farcs08$).}\label{a140_co10_hi.fig}
\end{figure}

In Fig.~\ref{a140_co10_hi.fig} we compare the distribution of \CO{1}{0} and
{\HI} associated with NGC~275. In the outer region of the galaxy the
co-incidence between the molecular gas peaks and the atomic gas peaks is good,
however the central molecular gas peak is offset to the north-east of the
atomic gas peak. As already noted, the {\HI} emission is offset towards the
south-west of the galaxy, whereas the CO data is well aligned with the optical
major axis of the system. Comparison of the kinematics of the {\HI} and CO data
was made by regridding the fully sampled data to a grid with $2''$ pixels and a
velocity resolution of 20 $\kms$ and smoothing the {\HI} data to the same
resolution of the CO dataset.

\begin{figure*}
\begin{tabular}{ccc}
\centering
\includegraphics[clip=,angle=0,width=6.5cm, bb=42 175 552 667, clip=]{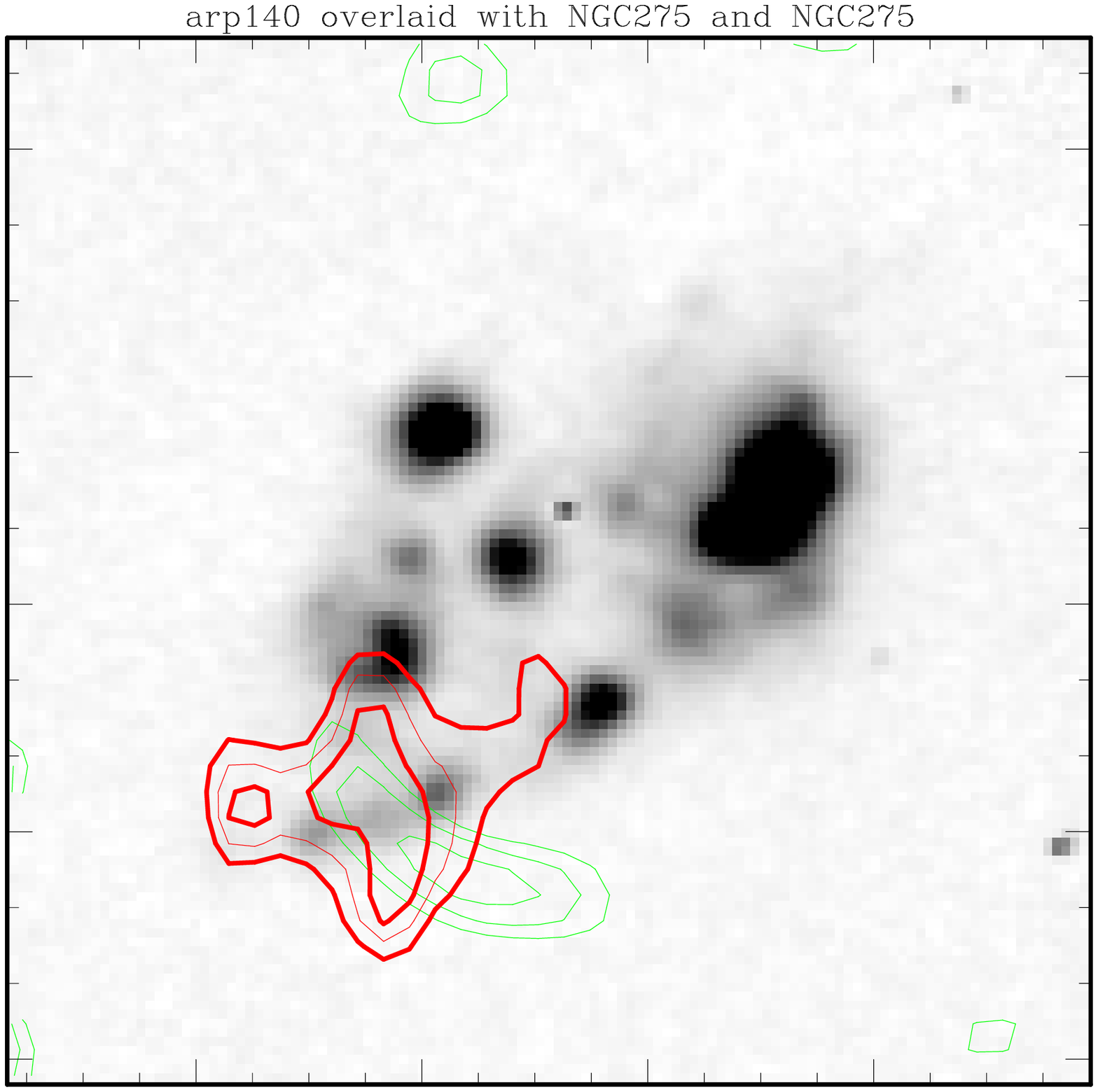} & &
    \includegraphics[clip=,angle=0, width=6.5cm, bb=42 175 552 668, clip=]{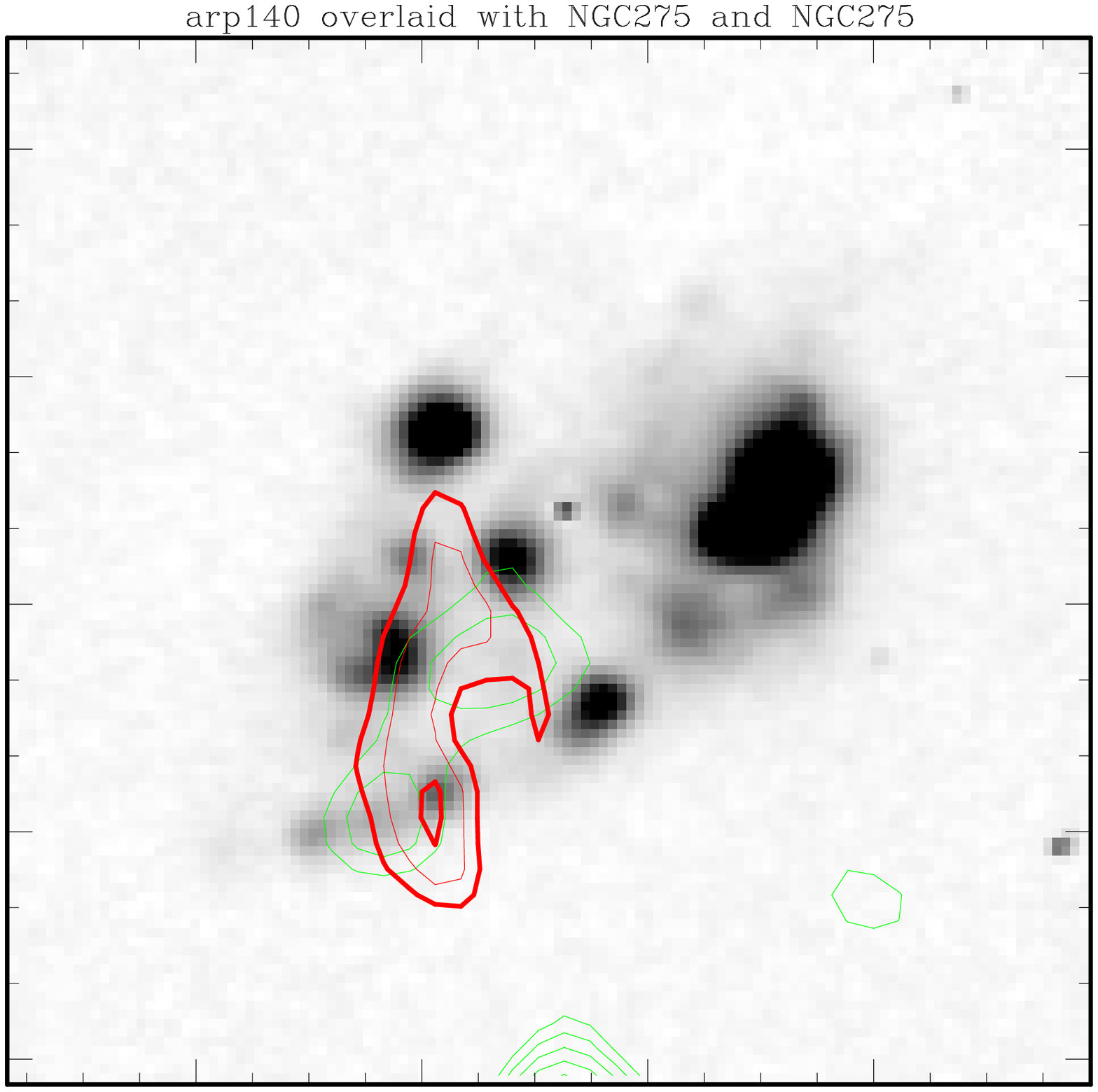} \\
    1733 $\kms$      & &   1754 $\kms$          \\ [3mm]
\includegraphics[clip=,angle=0,width=6.5cm, bb=42 175 552 667, clip=]{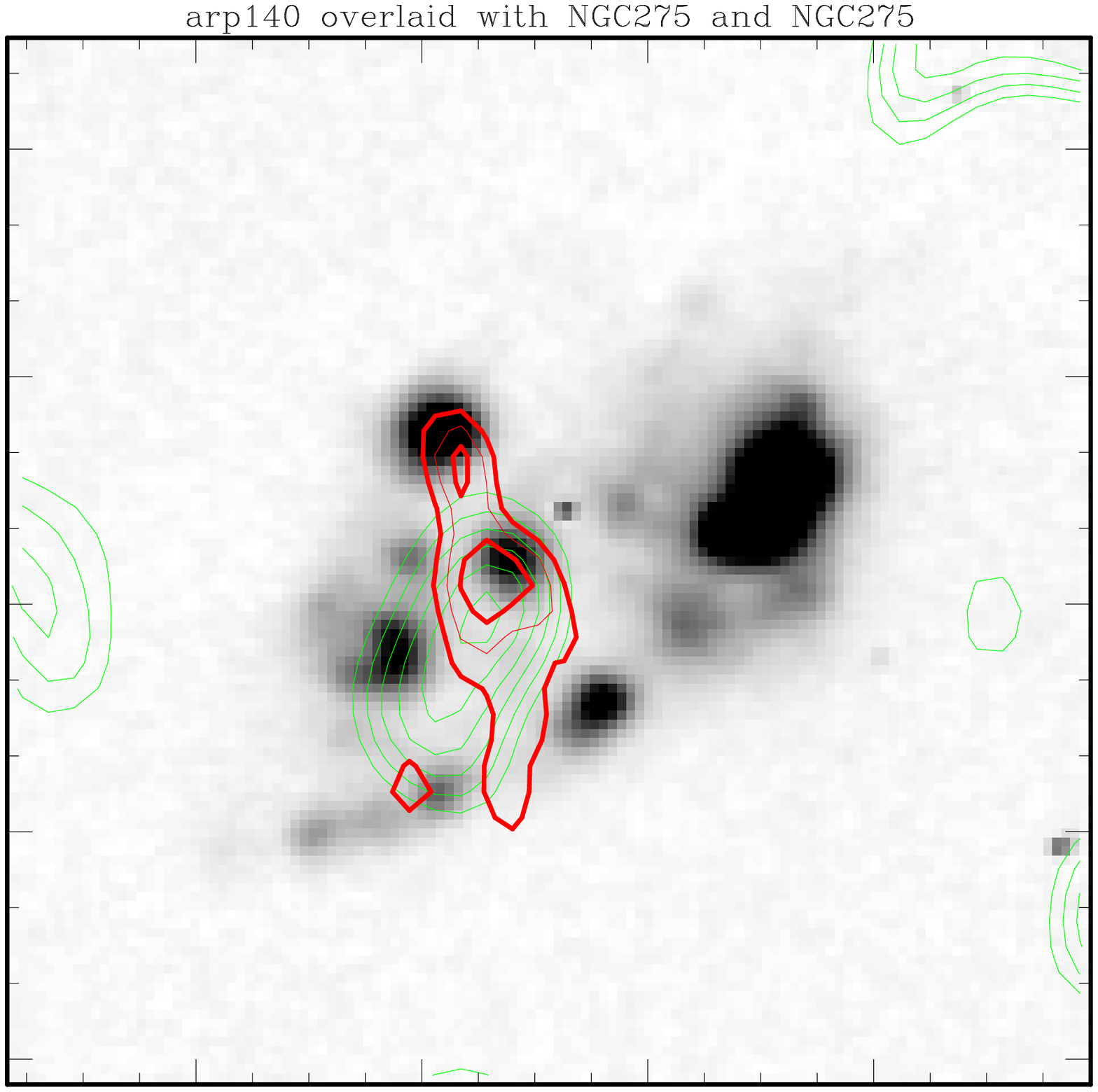} & &
    \includegraphics[clip=,angle=0, width=6.5cm, bb=42 175 552 668, clip=]{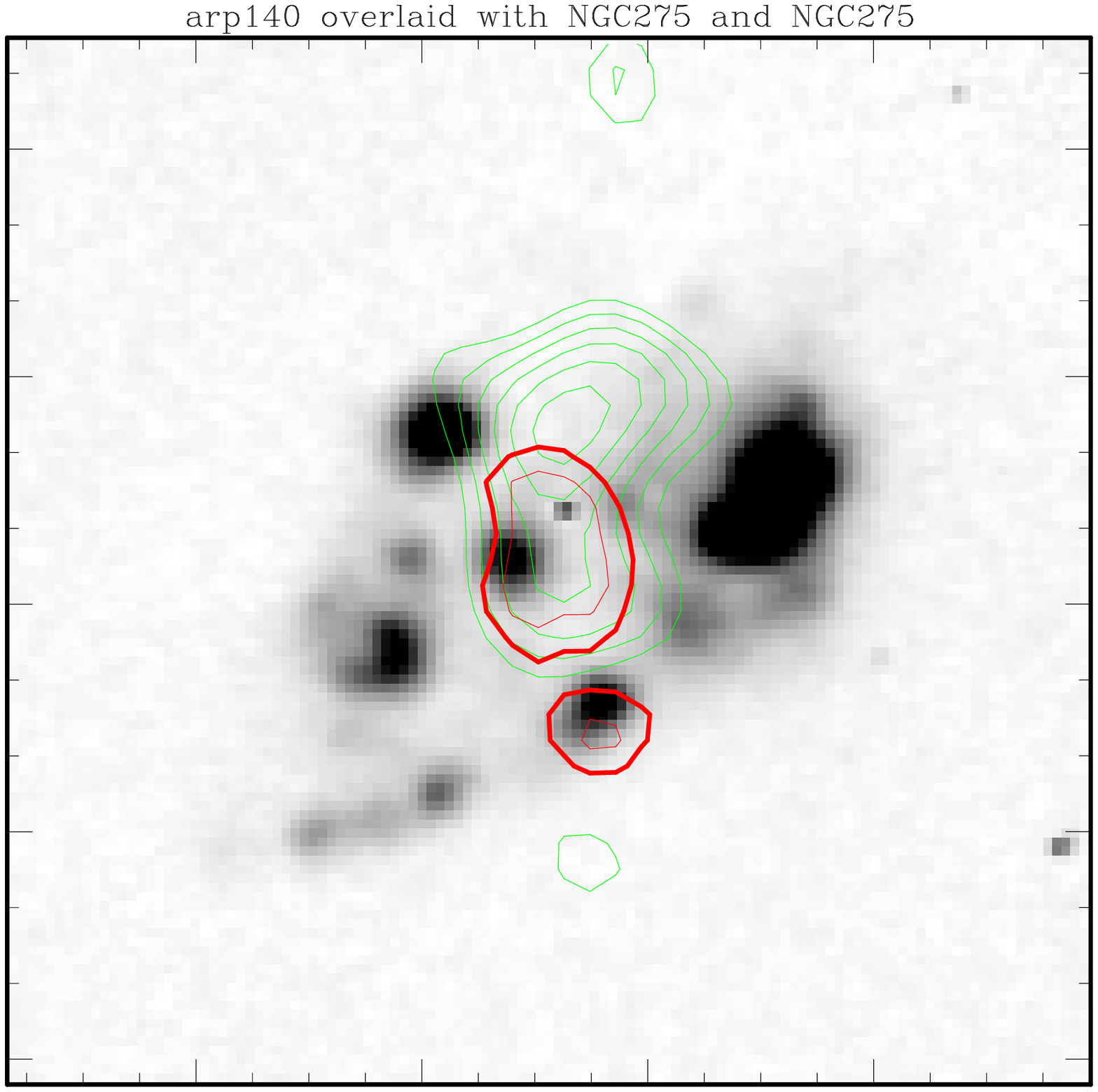} \\
    1775 $\kms$     & &  1796 $\kms$     \\ [3mm]
\includegraphics[clip=,angle=0,width=6.5cm, bb=42 175 552 667, clip=]{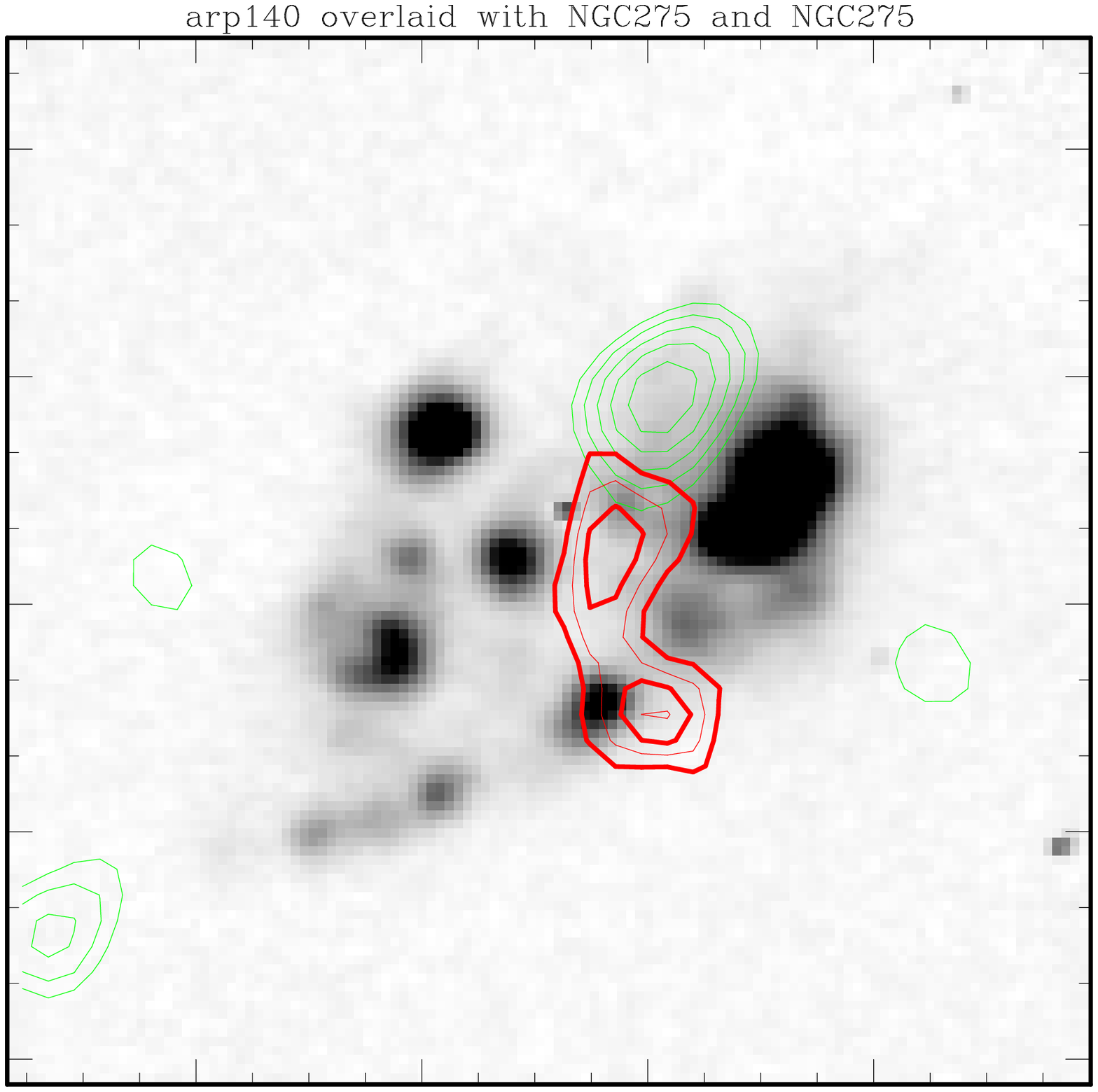} & &
   \\
    1817 $\kms$  & &       \\
\end{tabular}
\caption{Channel maps for the VLA B-array {\HI} (red contours) and combined
BIMA and OVRO \CO{1}{0} data (green contours), overlayed on greyscale of a
continuum subtracted H$\alpha$ image. Velocities indicated are the central
velocities of the 20 $\kms$ channels.}\label{channelmaps.fig}
\end{figure*}

Some channel maps of both {HI} and \CO{1}{0} emission are shown in
Fig.~\ref{channelmaps.fig}. The atomic and molecular gas appear spatially
co-incident at lower velocities, tracing emission in the south-eastern region
of the galaxy. This is also the region in which the molecular line-ratio
indicates the largest molecular gas densities. At the highest velocities,
tracing emission in the more northern region of the galaxy closer to the bright
{\HII} region, the spatial co-incidence breaks down. Spatial offsets between
the molecular and atomic gas, and a possible small kinematical offset occur in
the region of NGC~275 closest to the perturbing galaxy NGC~274.

\begin{figure}
\centerline{\includegraphics[clip=,angle=90,width=8.5cm]{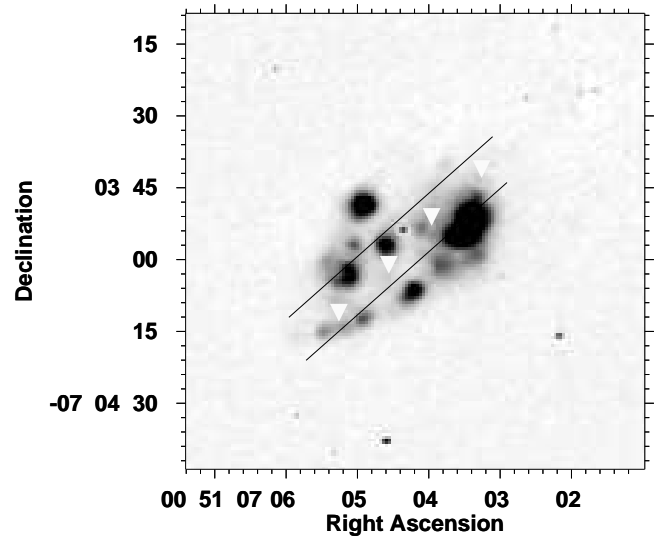}}
\caption{Triangles mark the positions for comparison of line intensity ratio of
\CO{2}{1} and \CO{1}{0} data. Lines mark the profile cuts for which we examine
the relative intensity of the two molecular lines.}\label{n275_ratiopts.fig}
\end{figure}

\begin{table}
\caption{The ratio of the \CO{2}{1} to \CO{1}{0} emission at four points
in NGC~275.}\label{275co_ratios}
\tabcolsep=4pt
\begin{tabular}{@{}cccccc} \hline
RA & Dec & $I_{10}$ &$I_{21}$ & $R_{21}$ & $\sigma_{R_{21}}$ \\
($^{\rm h}~^{\rm m}~^{\rm s}$) & (${^\circ}~'~''$) & (K~km~s$^{-1}$) & (K~km~s$^{-1}$) & & \\ \hline
00 51 05.3  & $-07$ 04 11.7  & 5.13  & 4.65  & 0.9  & 0.2  \\
00 51 04.6  & $-07$ 04 01.7  & 6.28  & 4.04  & 0.6  & 0.1  \\
00 51 04.0  & $-07$ 03 51.7  & 7.68  & 4.05  & 0.5  & 0.1  \\
00 51 03.3  & $-07$ 03 41.7  & 5.53  & 3.04  & 0.5  & 0.1  \\ \hline
\end{tabular}
\end{table}

\begin{figure*}
\centerline{\includegraphics[clip=,angle=270,width=85mm]{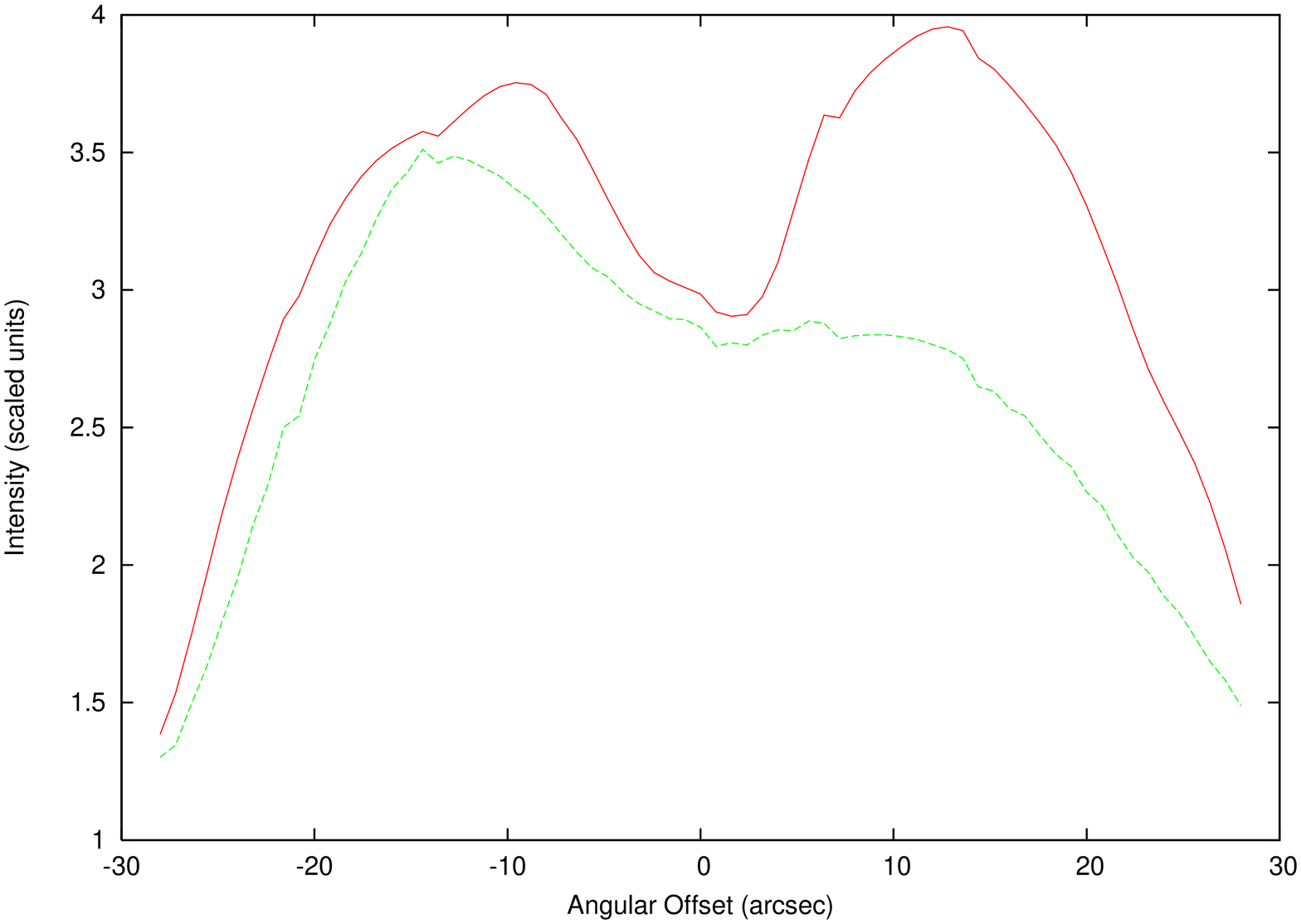}
            \qquad
            \includegraphics[clip=,angle=270,width=85mm]{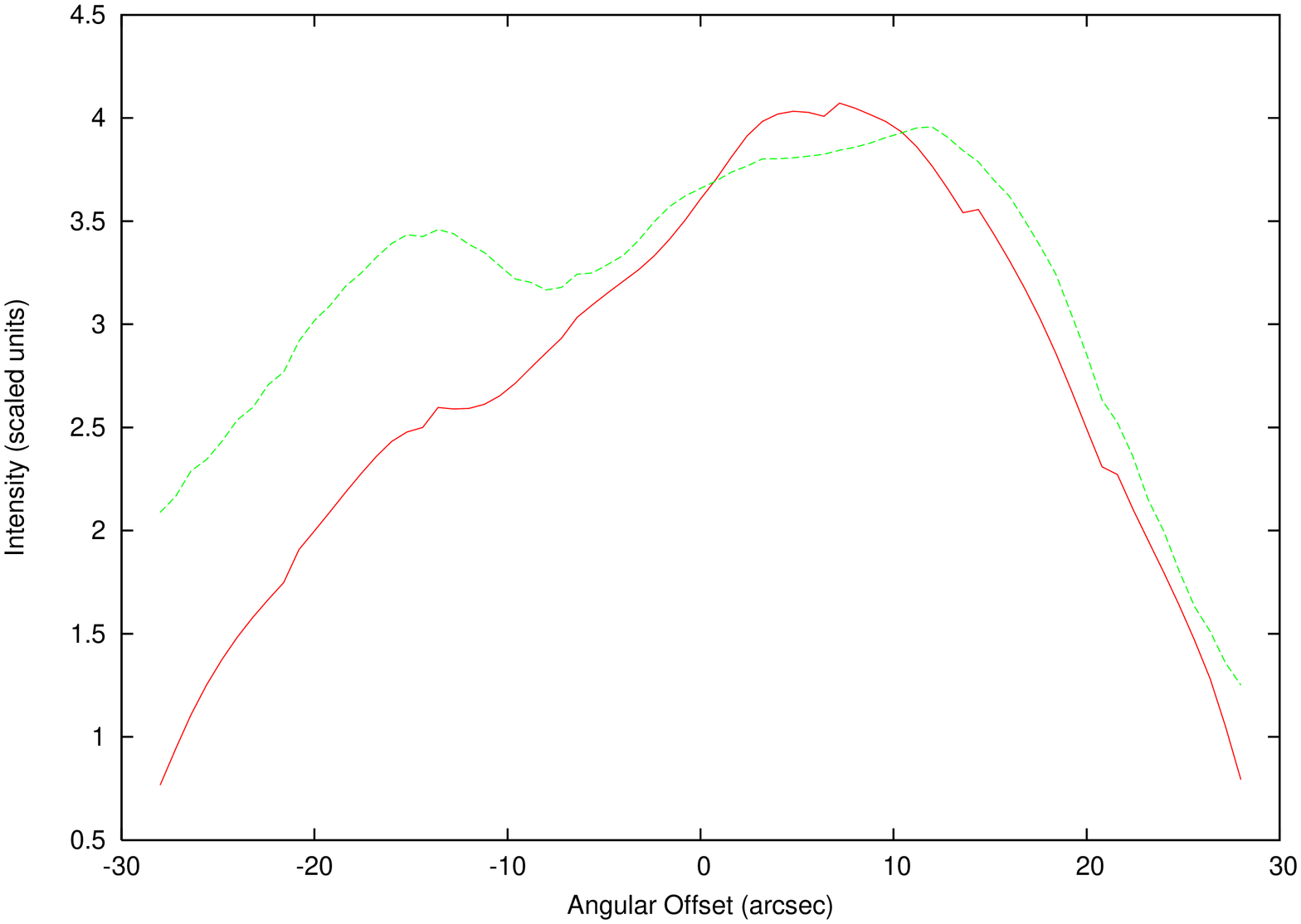}}
\caption{Two profile cuts comparing the relative intensity of the \CO{2}{1} and
\CO{1}{0} emission. The lower declination cut is displayed on the left and the
higher declination cut on the right.}\label{co_profile.fig}
\end{figure*}

To determine the \CO{2}{1}/\CO{1}{0} line ratio the \CO{1}{0} data were
smoothed to a resolution of $20''$, comparable to the JCMT data, and
re-gridded. The line ratio at four locations shown in
Fig.~\ref{n275_ratiopts.fig} are given in Table \ref{275co_ratios} and in
Fig.~\ref{co_profile.fig} we show profiles of the $J=2{-}1$ and $J=1{-}0$
emission along the two directions indicated in Fig.~\ref{n275_ratiopts.fig}.
The integrated line ratio for NGC~275 is $0.5\pm0.1$. The ratios in
Table~\ref{275co_ratios} indicate a trend of decreasing ratio as we move along
the major axis from east to west; the ratio decreases from $\sim0.9$ at the
south eastern end of NGC~275 to $\sim0.5$ in the north-western half of the
galaxy.  In the far south-eastern tip of NGC~275 the line ratio of $\sim 0.9$
is consistent with the properties of giant molecular clouds seen in Orion,
which have a kinetic temperature, $T_{k} \sim 20$ K and density $n({\rm H}_{2})
\geq 10^{3}$ cm$^{-3}$ \citep{1994ApJ...425..641S}. Away from the south-eastern
tip the line ratio falls to around $\sim0.6$. Assuming a typical value for the
kinetic temperature in the molecular clouds ($T_{k} > 10$ K), the decreasing
line ratio is consistent with a lower density of below $1 \times 10^{3}$
cm$^{-3}$ \citep{1994ApJ...425..641S}.

These results are interesting since they imply the densest hottest gas in this
system is not associated with the region of most active star formation. In
fact, the most active star forming region has the lowest line ratios. The
highest line ratios are to be found in the south-east of the galaxy, the region
suffering most disturbance as a result of the tidal interaction.  The line
ratio towards the centre of NGC~275 is significantly less than that found by
\cite{1992A&A...264..433B} in interacting systems.

\section{Relationship between star formation and gas distribution}

We now consider the origin of the strong anti-correlation between the
distribution of star-forming regions and molecular gas which we
see in NGC 275. The form of the molecular gas distribution is clearly
bar-like. NGC~275 is classified as a barred galaxy, and careful examination of
the optical and IR images reveal an indistinct bar which runs along the minor
axis. The CO morphology is disrupted, and the CO emission peaks at the edges of
the disk, as well as centrally.

\begin{figure}
\centerline{\includegraphics[clip=,angle=90,width=8cm]{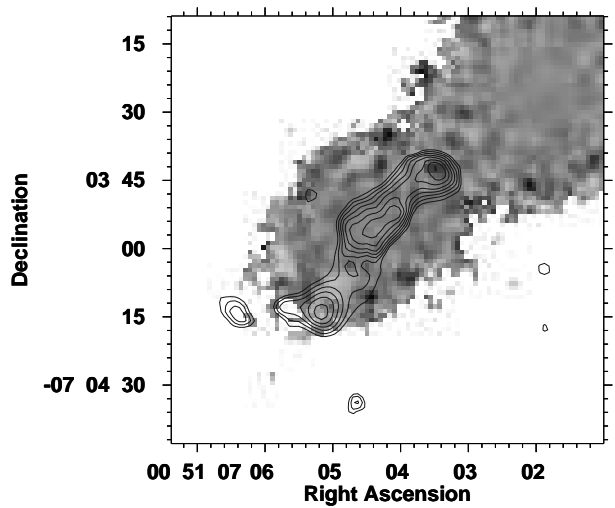} }
\caption{2MASS $J-H$ image overlayed with \CO{1}{0} combined BIMA and OVRO data
($\theta_{\rm FWHM} = 10\farcs30 \times 7\farcs63$). Contour levels correspond
to (2.0, 2.5, 3.0, 3.5, 4.0, 4.5, 5.0, 5.5) Jy beam$^{-1}$ km~s$^{-1}$. The
image is negative; light areas indicate small ratios corresponding to regions
which are more reddened. }\label{colour_colour1.fig}
\end{figure}

\begin{figure*}
\centerline{\includegraphics[clip=,origin=c, clip=, angle=90,width=60mm]{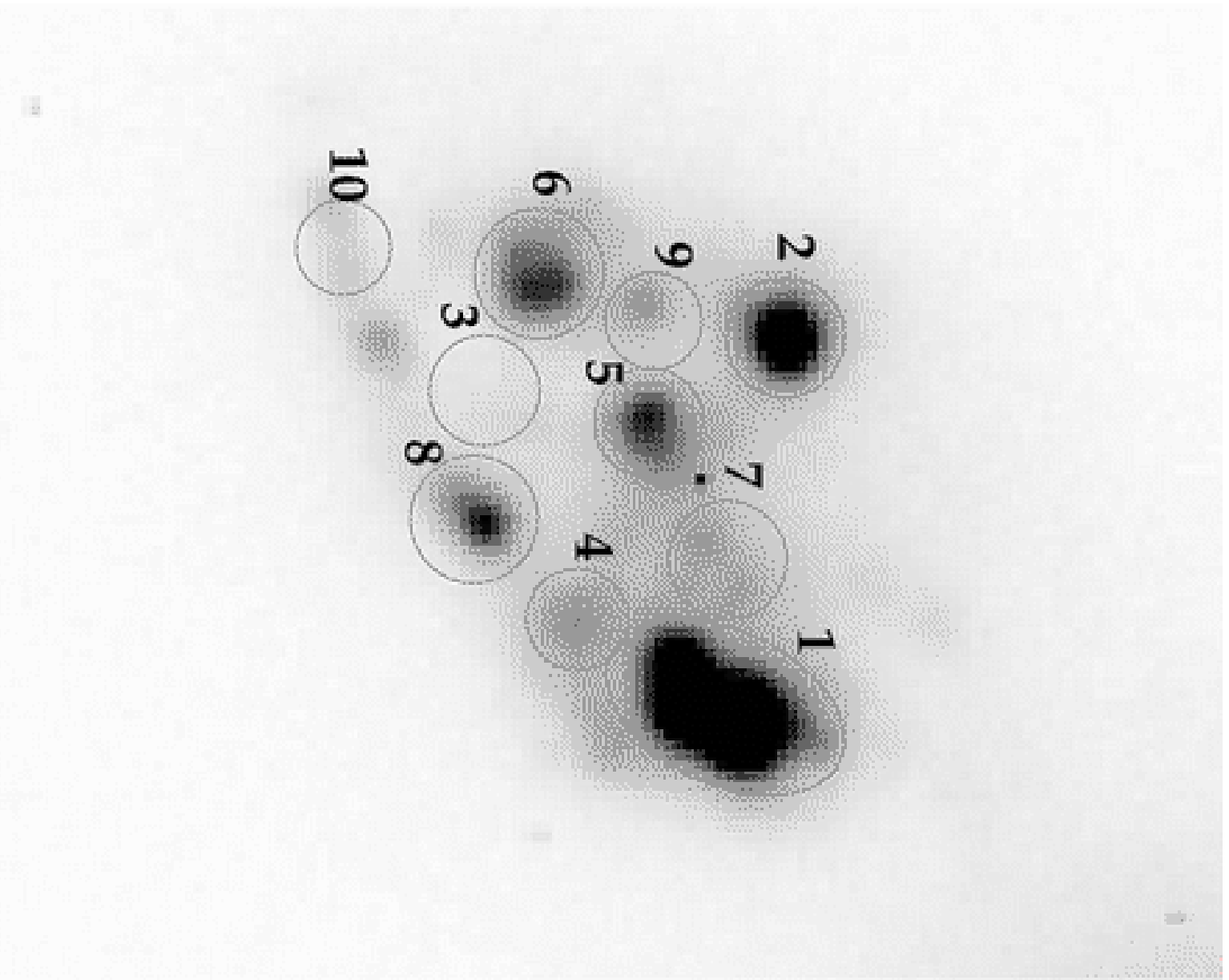}
            \qquad
            \includegraphics[clip=,origin=c, clip=,  angle=90,width=60mm]{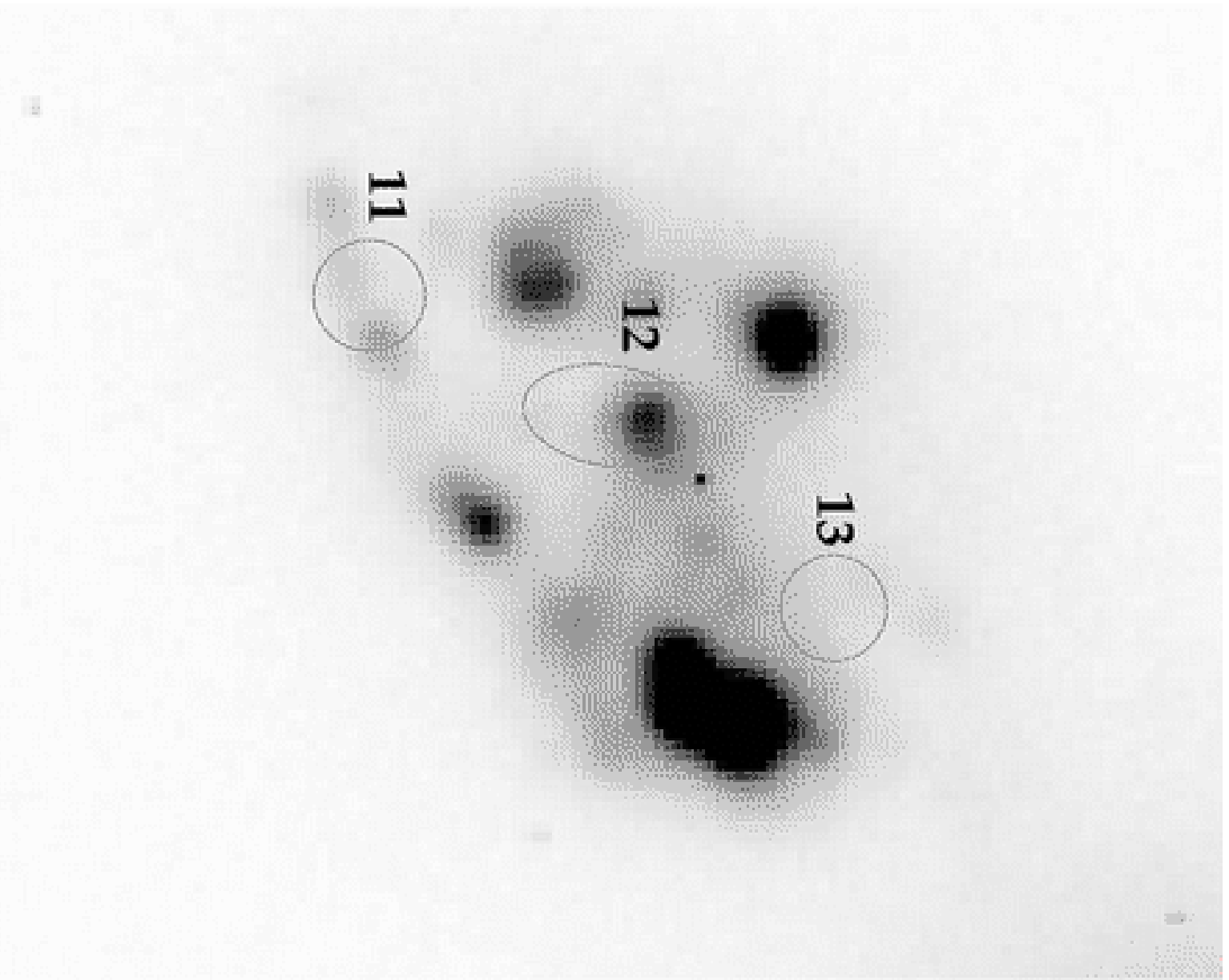}}
\caption{Images showing the regions where the H$\alpha$ predicted radio thermal
flux, total radio flux, gas column density and $J-H$ reddening were compared.
(left) Ten regions selected on the basis of the H$\alpha$ distribution. (right)
Three regions selected exhibiting high molecular gas column
densities.}\label{regions.fig}
\end{figure*}

First, we note that the large scale effects of SNe would be to
sweep up and/or ionise the neutral material, and hence produce a separation
between H$\alpha$ emission and the atomic gas. However, this is not seen. To
explain the anti-correlation between the distribution of molecular gas and the
sites of current star formation, we begin by examining the hypothesis that
extinction caused by dust associated with the molecular gas has given rise to
the apparent anti-correlation. Science grade images of NGC~275 in $J$, $H$ and
$K_{\rm s}$ bands were downloaded from the 2MASS archive \footnote{Available at
{\tt http://irsa.ipac.caltech.edu/applications/2MASS/LGA/}}. Colour difference
images of $J-H$, $H-K$ and $J-K$ were produced using {\sc IRAF} by blanking all
pixels below $3\sigma$ in the $J$- and $H$-images and below $2.5\sigma$ in the
$K_{\rm s}$-band image. The resulting $J-H$ image overlayed with contours
indicating the molecular gas distribution is shown in
Fig.~\ref{colour_colour1.fig}.  A quantitative analysis is given for several
regions, indicated in Fig.~\ref{regions.fig}, with the results shown in
Table~\ref{ha_radiocomp.tab}.  The largest reddening is associated with the
south-easterly peak of the molecular gas distribution which also has the
largest \CO{2}{1}/\CO{1}{0} ratio, consistent with a high gas density. There is
no evidence for significant reddening associated with the broad distribution of
molecular gas.  For all points along the molecular gas distribution the
reddening never exceeds 0.46 in $J-H$ and is typically only 0.2 magnitudes
larger than the observed reddening against peaks in the H$\alpha$ distribution.
For a standard extinction law this corresponds to much less than one magnitude
of extinction in H$\alpha$ at the worst.  We conclude that extinction caused by
dust associated with the molecular gas cannot account for the observed
distribution of H$\alpha$ emission. This conclusion is supported by the radio
continuum data which follow well the H$\alpha$ distribution and do not indicate
a region of hidden star formation behind the molecular gas.

\begin{table}
\caption{Extinction in NGC~275.}\label{ha_radiocomp.tab}
\begin{tabular}{@{}ccccccccc} \hline
Region & Size       & $S({\rm H}\alpha)10^{-14}$  & $J-H$ \\
       & kpc$^{2}$  & erg cm$^{-2}$ s$^{-1}$      & mag   \\
  (1)  & (2)        & (3)                         & (4)   \\ \hline
 1  & 1.54    & 45.27  & 0.06 \\
 2  & 0.57    & 8.36   & 0.08 \\
 3  & 0.48    & 1.59   & 0.30 \\
 4  & 0.40    & 3.51   & 0.11 \\
 5  & 0.48    & 4.18   & 0.15 \\
 6  & 0.67    & 6.66   & 0.10 \\
 7  & 0.55    & 3.94   & 0.16 \\
 8  & 0.64    & 4.95   & 0.04 \\
 9  & 0.38    & 2.37   & 0.12 \\
10  & 0.37    & 1.38   & 0.46 \\
11  & 0.49    & 2.18   & 0.37 \\
12  & 0.68    & 4.94   & 0.13 \\
13  & 0.44    & 1.99   & 0.11 \\ \hline
\end{tabular}\\
\medskip
(1) Region number: regions are shown in Fig.~\ref{regions.fig}.
(2) Area of region in kpc$^{2}$.
(3) The H$\alpha$ flux for region in question.
(4) The reddening observed in magnitudes using the 2MASS $J-H$ colour
difference image.
\end{table}

A second possibility is that molecular hydrogen has been dissociated by UV
photons from the regions of current star formation and/or molecular gas is
dissociated and ionised.  There is no evidence for large-scale dissociation of
molecular hydrogen to atomic hydrogen in the vicinity of the {\HII} regions ---
if this were to occur we would expect an anti-correlation between the atomic
and molecular gas, however we see no such effect when comparing the {\HI} and
\CO{1}{0} images. We can also rule out substantial ionisation.  Modelling the
observed H$\alpha$ peaks as individual {\HII} regions we can estimate an upper
limit to the column density of ionised hydrogen which we find to be typically
$5 \times 10^{18}$ cm$^{-2}$, which is more than three orders of magnitude less
than the observed column density of atomic and molecular gas towards the peaks
in the molecular gas distribution ($\sim 8 \times 10^{21}$ cm$^{-2}$).

A third possibility that we explore is that on-going star formation has
depleted the molecular gas close to the {\HII} regions.  We estimate the mass
of gas consumed during star formation by taking the current observed star
formation rate which is 1.5 M$_{\odot}$ yr$^{-1}$, and estimating the timescale
over which  star formation has been occurring.  To do this we assume that star
formation was triggered at perigalacticon, at which point the tidal
perturbation would have been a maximum. To estimate the time since
we use the observed distribution of {\HI} in the tidal tail. The
length of semi-circular tidal tail extending to the north of the interacting
pair is $\sim 3\farcm6$, corresponding to a physical distance of $\sim 26$ kpc.
Assuming the atomic gas associated with NGC~275 was not significantly
accelerated during the encounter and has moved ballistically since it was
stripped, we can estimate its velocity from the rotational speed of the galaxy
and its orbital velocity in the system as $v_{\rm gas} = {\Delta V}/2 + v$,
where $v$ is the rotational velocity of the galaxy from which the tidal tail
originated and $\Delta V$ is the relative radial velocity of the two
interacting galaxies. The interaction between NGC~275 and NGC~274 appears to be
largely in the plane of the sky; both galaxies having the same systemic
velocity giving a relative radial velocity of zero. From the rotation curve we
estimate the rotational velocity of NGC~275 to be $\sim175$ $\kms$ (corrected
for the inclination angle). This spatial extent of the tidal tail and the
estimated velocity of the gas give an approximate timescale of interaction of
$1.5\times10^{8}$ yr. Simple $N$-body simulations for this system
\citep{HarrietPhD} give a similar timescale of $\sim1.1 \times 10^{8}$ yr since
perigalacticon. Together with the current star formation rate  this gives for
the mass of molecular gas consumed since perigalacticon $(1{-}4) \times 10^{8}$
M$_{\odot}$; the upper end of the range represents approximately half the
current molecular gas mass in NGC~275. It is therefore quite possible that the
structure we observe in which the H$\alpha$ and molecular gas are spatially
anti-correlated is the result of consumption of the molecular gas during an
episode of star formation which began at about perigalacticon.

Another possible explanation for the observed distribution of molecular gas is
that it is a warped molecular disc or ring. NGC~275 is inclined with respect to
the plane of the sky (inclination angle $\sim 36.8$ degrees) and the systemic
velocities of the interacting galaxies are the same. These facts suggest that
the spin angular momentum vector of NGC~275 and the orbital angular momentum
vector of the interacting pair are not aligned, conditions consistent with the
production of a galactic warp.  The molecular gas in NGC~275 has an S-shaped
spatial distribution, which could be indicative of a warped disk.  It is not
unreasonable to postulate that the gas and stellar populations (and hence
{\HII} regions) have responded differently to  the tidal perturbation hence
giving rise to an offset between the gas distribution and the stellar disc.
However warps are usually observed on larger physical scales than that of the
molecular gas distribution in NGC~275 although a smaller scale warp was
recently found in the molecular gas distribution of NGC~3718
\citep{2004A&A...415...27P}; a system with a companion at 13 Mpc. There is
however no evidence for a warp in the {\HI} distribution; it therefore seems
unlikely that this mechanism could explain the observed distribution of gas.

\section{Star Formation in NGC~275}

NGC~275 is at a relatively advanced stage of interaction with an early-type
system, NGC~274. Estimates of the time since perigalacticon indicate a
timescale of approximately $1.5 \times 10^{8}$ yr. The galaxy has a relatively
normal SFR ($\sim 1.5$ M$_{\odot}$ yr$^{-1}$), although normalisation of the
modest FIR luminosity, $L_{\rm FIR}$, by an equally modest molecular gas yields
a sizeable star-formation efficiency ($L_{\rm FIR}/M_{\rm H_{2}} = 7.5$
L$_{\odot}$ M$_{\odot}^{-1}$). The distribution of molecular gas and regions of
star formation are unusual in a number of ways. Neither the molecular gas
distribution nor the recent star formation are centrally condensed, although
some star formation is occurring in the nucleus indicated by the presence of a
nuclear star cluster \citep*{2003AJ....125.1073B}. The star formation is patchy
and extended, and there is a strong anti-correlation with the molecular gas.
There is no evidence for a gravitational or dynamical origin of the observed
structures -- there is no indication of spiral structure in any data set, and
indeed the 2MASS $K_{\rm s}$-band data also show a patchy structure. Although
NGC~275 is a barred system, the bar is not strong, nor is the molecular gas
aligned with it. These observations and the analysis of the previous section
suggests that the anti-correlation we observe between the sites of current star
formation and the molecular gas distribution is most likely explained by the
consumption of gas during an episode of star formation which began close to
perigalacticon.

In galaxies in which large-scale structure such as spiral
arms organise the star formation systematic offsets are seen between the
molecular gas distribution and that of {\HII} regions
\citep{1988Natur.334..402V, 1993ApJ...404..593R, 1996MNRAS.283..251K,
1996ApJ...469L.101L}. These offsets are explained by the time delay between the
gas accumulation and onset of star formation. However, in galaxies which lack
such a large-scale driver a more random distribution of offsets are expected.
In the late-type galaxy M33, where spiral structure is weak,
\cite{2003ApJS..149..343E} and \cite{1991ApJ...370..184W} find that {\HII}
regions are often close to giant molecular clouds, but that there are many
examples of both molecular clouds with no {\HII} regions and {\HII} regions
without molecular clouds. Typically there is some offset between the two but
the sense of this seems not to be related to the spiral structure. The
situation in NGC 275 is probably closer to that in M33 than in galaxies with
strong spiral structure. The process of star formation in NGC 275 may have
followed a more stochastic evolution, with small-scale processes, rather than
galaxy-wide perturbations, dominating its evolution.

The lack of correlation between the molecular gas
distribution and that of star formation seen in NGC 275 may be a common
phenomenon. \cite{2005ApJS..158....1I} present {\HI} and interferometric CO
observations of 10 comparable-mass interacting galaxies, detecting isolated CO
emission with no optical counterpart in a number of systems. Similarly,
\cite{2001AJ....121..727W} observe no clear correlation between the {\HI} and
CO and sites of ongoing star formation in the dwarf starburst NGC 4214,
although the {\HI} and CO velocities agree well. The irregular nature of the
star formation in this galaxy, the absence of correlation between the total gas
column density and regions of star formation and the good agreement of the
{\HI} and CO dynamics is similar to the situation observed in NGC 275.

NGC~275's atomic and molecular gas content will be important in determining its
evolution. Consistent with other late-type spiral galaxies, NGC~275 has a
modest amount of molecular gas in a relatively extended distribution. The
atomic gas content of NGC~275 is similarly unremarkable, although its
morphology is more dramatic: the {\HI} having largely been stripped with only
one-third of the atomic gas still associated with the galaxy.

One of the defining features of the Arp 140 pair is that it contains only one
late-type, gas-rich galaxy. This characteristic distinguishes it from the
interacting systems that constitute the ULIRG population, believed to result
from the merger of two gas-rich disks. \cite{2004bdmh.confE..40M} are studying
a sample of mergers an order of magnitude less luminous then ULIRGs -- moderate
luminosity mergers with $L_{\rm FIR}$ luminosities in the range $\sim 5 \times
10^{9}{-}10^{11}$ L$_{\odot}$. They suggest that these systems might be the
product of a merger between two gas-rich galaxies of {\em unequal} mass or, as
is the case for Arp 140, a late-type and early-type galaxy, contrasting such
systems with ULIRGs, which are believed to result from the merger of two
comparable mass gas-rich disks, yielding very centrally condensed molecular gas
distributions and FIR luminosities in excess of $10^{12}$ L$_{\odot}$. The FIR
luminosity of NGC~275 places it at the faint end of the Manthey {et al.}
sample; although, NGC~275 is still a relatively young interaction when compared
with these merger systems.

For a star formation rate of 1.5~M$_{\odot}$ yr$^{-1}$ (the current rate in
NGC~275) the timescale for the consumption of the molecular gas is between
$(4.5{-}12)\times10^{8}$ yr. We estimate a time to merger of between 3 and $8
\times 10^{8}$ yr \citep{HarrietPhD}. As NGC~275 approaches merger most of its
supply of molecular gas will have been exhausted at the present rate of star
formation.  Unless there is significant conversion of atomic to molecular gas,
and also some infall of stripped material, there will not be a sufficient
supply of fuel to allow the star formation rate to increase significantly from
its present value.

\section{Conclusions}

Given the timescale required to produce the large {\HI} tidal
tail, NGC~275 is at a relatively advanced stage of interaction with an
early-type system. In contrast to the commonly painted picture of interacting
systems, the perturbation experienced by NGC~275 does not appear to have
triggered large-scale gas inflow and ensuing centrally enhanced star formation.
Despite NGC~275's interacting status and weakly barred potential, neither the
molecular gas distribution nor the recent star formation are centrally
condensed. The brightest emission from both the star-formation tracers and the
molecular gas are found not in the centre, but offset to the north-west. The
relationship between the molecular gas and H$\alpha$ emission in NGC~275 is
somewhat unusual; the two tracers appear to a large extent uncorrelated.
Investigation indicates this is unlikely to result from either extinction or
ionisation effects. Whilst there is some evidence of weak spiral structure in
this galaxy, the $K_{\rm s}$-band image does not indicate an underlying spiral
structure in NGC~275's mass distribution. Instead, the patchy H$\alpha$
emission is mirrored by equally patchy $K_{\rm s}$-band emission
(Section~\ref{275_star_form_rate.sec}). NGC~275 appears to be undergoing star
formation of a stochastic-like nature, in the sense that it is
driven by smaller scale feedback effects, rather than large scale dynamical
processes. This star formation, which may have been enhanced by the
interaction, is likely to have contributed to the depletion of molecular gas in
certain regions of the galaxy, shaping the observed gas distribution.

\section*{Acknowledgements}

We thank the staff of the OVRO and BIMA interferometers. The James Clerk
Maxwell Telescope is operated by The Joint Astronomy Centre on behalf of the
Particle Physics and Astronomy Research Council of the United Kingdom, the
Netherlands Organisation for Scientific Research, and the National Research
Council of Canada. The National Radio Astronomy Observatory is a facility of
the National Science Foundation operated under cooperative agreement by
Associated Universities, Inc. This research has made use of the NASA/IPAC
Extragalactic Database (NED) which is operated by the Jet Propulsion
Laboratory, California Institute of Technology, under contract with the
National Aeronautics and Space Administration. HC acknowledges receipt of a
PPARC studentship.


\bigskip

\label{lastpage}
\end{document}